\documentclass[11pt]{article}
\usepackage{epsfig}

\begin{document}

\newcommand{\bm}[1]{\mbox{\boldmath$#1$}}
\newcommand{\unbfm}[1]{\mbox{\boldmath$#1$}}

\def\etal{{\it et al }}      
\def\kmh#1#2{\underline{#1} \{#2\}}  
\def\given{\,|\,}   
\def\intd{\,\mbox{d}}    
\def\mvec#1{{\bm{#1}}}   
%

\title{Bayesian Inference in Processing Experimental Data \\
Principles and Basic Applications}
\author{G.~D'Agostini \\
Universit\`a ``La Sapienza'' and INFN, Roma, Italia
}
\date{}

\maketitle

\begin{abstract}
This report introduces general ideas and some
basic methods of the Bayesian probability theory
applied to physics measurements.
Our aim is to make the reader familiar, through examples
rather than rigorous formalism, with concepts such as:
model comparison
(including the automatic {\it Ockham's Razor} filter provided
by the Bayesian approach); parametric inference; 
quantification of the uncertainty
about the value of physical quantities, also taking 
into account systematic effects; role of marginalization;
posterior characterization; 
predictive distributions; 
hierarchical modelling and hyperparameters;
Gaussian approximation of the posterior 
and recovery of conventional methods, especially maximum likelihood
and chi-square fits under well defined conditions;
conjugate priors, transformation invariance and maximum
entropy motivated priors; Monte Carlo estimates of expectation, 
including a short introduction to Markov Chain Monte Carlo
methods.~\footnote[0]{Invited 
article to be published in {\it Reports on Progress in Physics}.

     {\rm Email}: {\tt dagostini@roma1.infn.it},
     {\rm URL}: {\tt http://www.roma1.infn.it/\,$\tilde{ }$\,dagos.}
}
\end{abstract}

\section{Introduction}
\label{sect:intro}
The last decades of 20$^{\rm th}$
century have seen an intense expansion in the use of Bayesian methods in
all fields of human activity that generally deal with
uncertainty, including
 engineering, computer science,
economics,
medicine  and even forensics (Kadane and Schum 1996). 
Bayesian networks (Pearl 1988, Cowell \etal 1999) are used to diagrammatically 
 represent uncertainty in expert systems or to construct artificial
 intelligence systems.
Even venerable metrological associations, such as 
the International Organization for Standardization
(ISO 1993), the Deutsches Institut f\"ur Normung (DIN 1996, 1999), 
and the USA National Institute of Standards and 
Technology (Taylor and Kuyatt 1994), have come to
realize that Bayesian ideas are essential to
provide general methods for quantifying uncertainty in
measurements. A short account of the Bayesian upsurge can
be found in a Science article (Malakoff 1999).  A search
on the web for the keywords `Bayesian,' `Bayesian network,' or
`belief network' gives one a dramatic impression of this
`revolution,' not only in terms of improved methods, but more importantly 
in terms of  reasoning. 
An overview of recent developments in Bayesian statistics, may be found in the proceedings
of the Valencia Conference series. 
The last published volume was (Bernardo \etal 1999), and the
most recent conference was held in June 2002. Another series of workshops, 
under the title of {\it Maximum Entropy and Bayesian Methods}, 
 has focused more on applications in the physical sciences.

It is surprising that many physicists
have  been slow to adopt
these `new' ideas. There have been notable exceptions, of course, 
many of whom have contributed to the abovementioned
Maximum Entropy workshops. One reason to be surprised is
because numerous great physicists and mathematicians have played
important roles in developing probability theory. 
These `new' ideas actually originated long ago with Bernoulli, Laplace, and
Gauss, just to mention a few who contributed significantly 
to the development of physics, as well as to Bayesian thinking. 
So, while modern statisticians and mathematicians
are developing powerful methods to apply to Bayesian analysis, 
most physicists, in their use
and reasoning in statistics still rely on
20$^{\rm th}$ century `frequentist prescriptions' (D'Agostini 1999a, 2000).

We hope that this report will help fill this gap by reviewing the 
advantages of using the Bayesian approach to address physics problems. 
We will emphasize more the intuitive and
practical aspects than the theoretical ones. We will not try to
cover all possible applications of Bayesian analysis in physics, but mainly
concentrate on some basic applications that illustrate clearly the
power of the method and how naturally it meshes with physicists' 
approach to their science.  

The vocabulary, expressions, and examples have been chosen with the intent
to correspond, as closely as possible, to the education that
 physicists receive in
statistics, instead of a more rigorous approach that 
formal Bayesian statisticians might prefer. For example, we avoid 
many important theoretical concepts, like exchangeability, 
and do not attempt to prove the basic rules of probability. When we talk 
about `random variables,' we will
in fact mean `uncertain variables,' and instead of referring
to the frequentist concept of `randomness'
\`a la von Mises (1957).  This distinction will be clarified later.

In the past, presentations on Bayesian probability theory often start
 with criticisms of `conventional,' that is,
frequentist ideas, methods, and results. We shall keep criticisms and 
detailed comparisons of the
results of different methods to a minimum. 
Readers interested in a critical review of
conventional frequentist statistics will find a large literature, 
because most introductory books or reports on Bayesian analysis 
contain enough material on this matter. 
See (Gelman \etal 1995, Sivia 1997, D'Agostini 1999c, 
Jaynes 1998, Loredo 1990)
and the references therein. 
Eloquent `defenses of the Bayesian choice' can be found in 
(Howson and Urbach 1993, Robert 2001).

Some readers may wish to have references to unbiased comparisons
of frequentist to Bayesian ideas and methods.
To our knowledge, no such reports exist. Those
who claim to be impartial are often
frequentists who take some Bayesian results as if
they were frequentist `prescriptions,'
not caring whether all underlying hypotheses apply.
For two prominent papers of this kind,
see the articles by Efron (1986a) [with follow up discussions by Lindley (1989),
 Zellner (1986), and Efron (1986b)] 
and Cousins (1995). A recent, pragmatic comparisons of 
frequentist and Bayesian confidence limits can be found in (Zech 2002). 

Despite its lack of wide-spread use in physics,
and its complete absence in physics courses (D'Agostini 1999a), 
Bayesian data analysis is 
increasingly being employed in many areas of physics, for example, 
in astronomy (Gregory and Loredo 1992, 1996, Gregory 1999, 
Babu and Feigelson 1992, 1997, Bontekoe \etal 1994), 
in geophysics (Glimm and Sharp 1999),
in high-energy physics (D'Agostini and Degrassi 1999, Ciuchini \etal 2001), 
in image reconstruction (Hanson 1993), 
in microscopy (Higdon and Yamamoto 2001),
in quantum Monte Carlo (Gubernatis \etal 1991), and 
in spectroscopy (Skilling 1992, Fischer \etal 1997, 1998, 2000), 
just to mention a few articles written in the last decade.  
Other examples will be cited throughout the paper.

\section{Uncertainty and probability}
\label{sect:probability}
In the practice of science, we constantly find ourselves in a state
of uncertainty. Uncertainty about the data that an experiment {\it shall} yield.
Uncertainty about the {\it true value} of a physical quantity,
even after an experiment has been done. Uncertainty about model parameters,
calibration constants, and other quantities that might influence the outcome
of the experiment, and hence influence our conclusions about the quantities of
interest, or the models that might have produced the observed results.

In general, we know through experience that not all the {\it events}
that could happen, or all conceivable {\it hypotheses}, are
equally likely. 
Let us consider the outcome of \underline{you} measuring 
the temperature at the location where you are presently reading
this paper, assuming you use a digital thermometer with one
degree resolution (or you round the reading at the degree if you have a
more precise instrument). 
There are some values
of the thermometer display you are more confident to read, others
you expect less, and extremes you do not believe at all (some of
them are simply excluded by the thermometer you are going to
use). Given two events $E_1$ and $E_2$, for example $E_1 :
\mbox{``}T = 22^\circ  \mbox{C''}$ and
 $E_2 : \mbox{``}T = 33 ^\circ \mbox{C''}$, you might consider
 $E_2$ much more {\it probable}  than $E_1$, just meaning
that you {\it believe} $E_2$ to happen more than $E_1$. We could
use different expressions to mean exactly the same thing:
you consider  $E_2$ more likely; you are more confident in $E_2$;
having to choose between  $E_1$ and $E_2$ to win a price, you
would promptly choose $E_2$; having to classify with a number, that
we shall denote with $P$, your degree of confidence on the two
outcomes, you would write  $P(E_2) > P(E_1)$; and many others.

On the other hand, we would rather state
the opposite, i.e. $P(E_1) > P(E_2)$, with the same meaning of
symbols and referring exactly to the same events:\ what
you are going to read at your place with your
thermometer. The reason is simply because we 
do not share the same
status of information. We do not know who you are and where you
are in this very moment. You and we are uncertain about the same
event, but in a different way. Values that might appear very probable
to you now, appear quite improbable, though not impossible, to us.

In this example we have introduced two crucial aspects of the
Bayesian approach:
\begin{enumerate}
\item
As it is used in everyday language, the term probability has 
the intuitive meaning of 
{\it ``the degree of belief that an event will occur.''}
\item
Probability depends on our state of knowledge, which is usually different
for different people. In other words, probability is 
unavoidably {\it subjective}. 
\end{enumerate}

At this point, you might find all of this quite
natural, and wonder why these intuitive concepts go by 
the esoteric name `Bayesian.' We agree! The fact is that the main thrust of statistics
theory and practice during the 20$^{\rm th}$ century has been based on a
different concept of probability, in which it is defined as the limit of the 
long-term relative frequency of the outcome of these events.  
It revolves around the theoretical notion of infinite
ensembles of `identical experiments.'
Without
entering an unavoidably long critical discussion of the frequentist approach,
we simply want to point out that in such a framework, there is
no way to introduce the probability of hypotheses. All practical methods
to overcome this deficiency yield misleading, and even
absurd, conclusions.
See (D'Agostini 1999c) for several examples and also for 
a justification of why frequentistic test `often work'. 

Instead, if we recover the intuitive concept
of probability, we are able to talk in a natural way about the
probability of any kind of event, or, extending the concept,
of any {\it proposition}. 
In particular, the probability evaluation based on the relative frequency of 
similar events occurred in the past is easily recovered in the 
Bayesian theory, under precise condition of validity
(see  Sect.~\ref{sect:binomial}).
Moreover, a simple theorem from
probability theory, Bayes' theorem, which we shall see in the next section,
allows us to update probabilities on the basis of new
information. This inferential use of Bayes' theorem is
only possible if probability is understood in terms of degree of belief.
Therefore, the terms `Bayesian' and `based on subjective probability'
are practically synonyms,and usually mean `in contrast to the
frequentist, or conventional, statistics.' The terms
`Bayesian' and `subjective' should be considered transitional.
In fact, there is already the tendency among many Bayesians 
to simply refer to `probabilistic
methods,' and so on (Jeffreys 1961, de Finetti 1974, Jaynes 1998
and Cowell \etal 1999).

As mentioned above, Bayes' theorem plays a fundamental role in
the probability theory. This means that subjective probabilities of logically
connected events are related to each other by mathematical rules. 
This important result can be summed up by saying, in
practical terms, that {\it `degrees of belief follow the same
grammar as abstract axiomatic probabilities.'} Hence, all formal
properties and theorems from probability theory follow.

Within the Bayesian school, there is no single way to derive 
the basic rules of probability (note that they are not 
simply taken as axioms in this approach). 
de Finetti's principle of {\it coherence} 
(de Finetti 1974) is considered
the best guidance by many leading Bayesians
(Bernardo and Smith 1994, O'Hagan 1994, Lad 1996 and
Coletti and Scozzafava 2002). 
See (D'Agostini 1999c) 
for an informal introduction to the concept of coherence, which in simple
words can be outlined as follows. A person who evaluates
probability values should be ready to accepts bets in either direction,
with odd ratios calculated from those values of probability. 
For example, an analyst that declares to be confident 50\% on $E$
should be aware that somebody could ask him to make a 1:1 bet 
on $E$ or on $\overline E$. If he/she feels uneasy, it means that 
he/she does not consider the two events equally likely and the
50\% was `incoherent.'

Others,
in particular practitioners close to the
Jaynes' Maximum Entropy school (Jaynes 1957a, 1957b)
 feel more at ease with  Cox's logical consistency reasoning, 
requiring some consistency properties (`desiderata')
between values of probability related to logically connected propositions.
 (Cox 1946).
 See also (Jaynes 1998, Sivia 1997, and  Fr\"ohner 2000,
and especially Tribus 1969),
 for accurate derivations and
a clear account of the meaning and
 role of information entropy in data analysis.
An approach similar to  Cox's is followed
by Jeffreys (1961), another leading figure
who has contributed a new vitality to the
methods based on this `new' point of view on
probability. Note that Cox and Jeffreys were
physicists.  Remarkably,
 Schr\"odinger (1947a, 1947b)
also arrived at similar conclusions, though his 
definition of event is closer to the de Finetti's one.
[Some short quotations from (Schr\"odinger 1947a) are
in order. Definition of probability: {\it ``\ldots a quantitative measure 
of the strength of our conjecture or anticipation, founded
on the said knowledge, that the event comes true''}. 
Subjective nature of probability: 
``{\it Since the knowledge may be different with different persons
or with the same person at different times, they may anticipate
the same event with more or less confidence, and thus different numerical
probabilities may be attached to the same event.''} Conditional probability:
{\it ``Thus whenever we speak loosely of `the probability of an event,'
it is always to be understood: probability with regard to a certain 
given state of knowledge.''}] 

\section{Rules of probability}
\label{sect:rules}

We begin by stating some basic rules and
general properties that form the `grammar' of the probabilistic
language, which is used in Bayesian analysis. In this section, 
we review the rules of probability, starting with the rules for 
simple propositions.  We will not provide rigorous derivations 
and will not address the foundational or philosophical
aspects of probability theory. Moreover,  following an `eclectic'
approach which is common among Bayesian practitioners, we talk
indifferently about probability of events, probability of hypotheses or 
probability of propositions. Indeed, the last expression will be often
favoured, understanding that it does include the others.

\subsection{Probability of simple propositions}
\label{sect:boolean}

Let us start by recalling the basic rules of probability for propositions or hypotheses. 
Let $A$ and $B$ be propositions,
which can take on only two values, for example, true or false.  
The notation $P(A)$ stands for the probability that $A$ is true. 
The elementary rules of probability for simple propositions are
\begin{eqnarray}
&& 0\le P(A) \le 1; \label{eq:basic1}\\
&& P(\Omega) = 1; \label{eq:basic2}\\
&& P(A\cup B) = P(A)+P(B) - P(A\cap B) \,. \label{eq:basic3} 
\\
&& P(A\cap B) = P(A\,|\,B) \,  P(B) = P(B\,|\,A) \,  P(A)\,,
  \label{eq:joint}
\end{eqnarray}
where $\Omega$ means {\it tautology} (a proposition that is 
certainly true).
 The construct $A\cap B$ is true only when both $A$ and $B$ are true (logical AND),
while  $A\cup B$ is true when at least one of the
two propositions is true (logical OR).  $A\cap B$ is also
written simply as `$A,B$' or $A B$, and is also called a {\it logical product},
while $A\cup B$ is also called a {\it logical sum}. 
$P(A, B)$ is called the joint probability of $A$ and $B$.
$P(A\,|\,B)$ is the probability of $A$ 
under that condition that $B$ is true. We often read it 
simply as ``the probability of A, given $B$.''  .

Equation (\ref{eq:joint})  shows that the joint probability of 
two events can be decomposed into conditional probabilities 
in different two ways. 
Either of these ways is called the {\it product rule}.
If the status of $B$ does not change the probability
of $A$, and the other way around, then
$A$ and $B$ are said to be {\it independent}, {\it probabilistically} 
independent to be precise. In that case, $P(A\,|\,B)=P(A)$, and
$P(B\,|\,A)=P(B)$, which, when inserted in Eq.~(\ref{eq:joint}), yields
\begin{eqnarray}
P(A\cap B) &=& P(A) \,  P(B)\hspace{.5cm}
\Longleftrightarrow\hspace{.5cm}\mbox{\it probabilistic independence} \, .
\label{eq:joint_ind}
\end{eqnarray}
Equations~(\ref{eq:basic1})--(\ref{eq:joint}) logically lead to
other rules which form the body of probability theory.
 For example,
indicating the {\it negation} (or {\it opposite}) 
of $A$ with $\overline{A}$, clearly $A \cup \overline{A}$
is a tautology ($A \cup \overline{A}=\Omega$),
and $A \cap \overline{A}$ is a contradiction
($A \cap \overline{A}=\emptyset$). 
The symbol $\emptyset$ stands for contradiction
(a proposition that is certainly false).
Hence, we obtain from Eqs.~(\ref{eq:basic2}) and (\ref{eq:basic3})
\begin{eqnarray}
P(A) + P(\overline{A}) &=& 1\, , 
\label{eq:normalization0} 
\end{eqnarray}
which says that proposition $A$ is either true or not true.

\subsection{Probability of complete classes}
\label{sect:boolean2}
These formulae become more interesting when we consider a set of 
propositions $H_j$ that all together form a tautology
(i.e., they are {\it exhaustive}) and are mutually {\it exclusive}.
Formally
\begin{eqnarray}
\cup_i H_j &=& \Omega \label{eq:cclass1} \\
H_j \cap H_k &=& \emptyset \quad \mbox{if }\ j \ne k \,
\,.
\label{eq:cclass2}
\end{eqnarray}
When these conditions apply, the set $\{H_j\}$
is said to form a {\it complete class}. 
The symbol $H$ has been chosen because we shall soon interpret 
$\{H_j\}$ as a set of {\it hypotheses}.

The first (trivial) property
of a complete class is {\it normalization}, that is
\begin{eqnarray}
\sum_j P(H_j) &=& 1\,, 
\label{eq:normalization}
\end{eqnarray}
which is just an extension of Eq.~(\ref{eq:normalization0}) 
to a complete class 
containing more than just a single proposition and its negation.

For the complete class $H$, the generalizations of
Eqs.~(\ref{eq:normalization0}) and the use of 
Eq.~(\ref{eq:joint}) yield:
\begin{eqnarray}
P(A) &=& \sum_j P(A, H_j) \label{eq:decomp4} \\
P(A) &=& \sum_j P(A\,|\,H_j) \,  P(H_j) \,. 
\label{eq:decomp5}
\end{eqnarray}
Equation~(\ref{eq:decomp4}) is the basis of what is called {\it
marginalization}, which will become particularly important when
dealing with uncertain variables:\ the probability of $A$ is
obtained by the summation over all possible 
{\it constituents}
 contained in $A$.  Hereafter, we avoid explicitly writing the
limits of the summations, meaning that they extend over all
elements of the class. The constituents are `$A, H_j$,' which,
based on the complete class of hypotheses $\{H\}$, themselves form
 a complete class, which can be easily proved.
 Equation~(\ref{eq:decomp5}) shows that the probability of any proposition
is given by a weighted average of all conditional probabilities,
subject to hypotheses $H_j$ forming a complete class, with the weight
being the probability of the hypothesis.

In general, there are many ways to choose complete classes 
(like `bases' in geometrical spaces). Let us denote the
elements of a second complete class by $E_i$. The constituents are then
formed by the elements $(E_{i},H_j)$ of the Cartesian product
$\{E\}\times\{H\}$. Equations~(\ref{eq:decomp4}) and
(\ref{eq:decomp5}) then  become the more general statements
\begin{eqnarray}
P(E_i) &=& \sum_j P(E_i, H_j) \label{eq:decomp6} \\
P(E_i) &=& \sum_j P(E_i\,|\,H_j) \,  P(H_j)  \label{eq:decomp7}
\end{eqnarray}
and, symmetrically,
\begin{eqnarray}
P(H_j) &=& \sum_i P(E_i, H_j) \label{eq:decomp8} \\
P(H_j) &=& \sum_i P(H_j\,|\,E_i) \,  P(E_i)\,.  \label{eq:decomp9}
\end{eqnarray}
The reason we write these formulae both ways is to
stress the symmetry of Bayesian reasoning with respect to
classes  $\{E\}$ and $\{H\}$, though we shall soon associate them with {\it observations} (or {\it events}) and {\it hypotheses}, respectively.

\subsection{Probability rules for uncertain variables}
\label{sect:uncertainty}

In analyzing the data from physics experiments, we need to deal 
with measurement that are discrete or continuous in nature. 
Our aim is to make inferences about the models that we believe 
appropriately describe the physical situation, 
and/or, within a given model, to determine the values 
of relevant physics quantities.
Thus, we need 
the probability rules that apply to uncertain variables, 
whether they are discrete or continuous. The rules for complete 
classes described in the preceding section clearly apply directly 
to discrete variables. With only slight changes, the same rules 
also apply to continuous variables because they may be thought 
of as a limit of discrete variables, as interval between 
possible discrete values goes to zero. 

For a discrete variable $x$, the expression $p(x)$, which is called 
 a {\it probability function}, has the interpretation in terms of the 
probability of the proposition
 $P(A)$, where $A$ is true when the value of the variable is equal to $x$. 
In the case of continuous variables, 
we use the same notation, but with the meaning of a {\it probability
density function} (pdf). So $p(x) \, dx$, in terms of a proposition,
is the probability $P(A)$, where $A$ is true when the value of the variable 
lies in the range of $x$ to $x + dx$.
In general, the meaning is
clear from the context; otherwise it should be stated.  
Probabilities involving more than one variable, like $p(x,y)$, 
have the meaning of the probability
of a logical product; they are usually called {\it joint}
probabilities.

Table \ref{tab:fx} summarizes useful formulae for discrete and
continuous variables. The interpretation and use of these relations 
in Bayesian inference will be illustrated in the following sections. 

\begin{center}
\begin{table}
\caption{\sl Some definitions and properties of 
probability functions for values 
of a discrete variable $x_i$ and
probability density functions for continuous variables 
$x$. All summations and integrals are understood to extend 
over the full range of possibilities of the variable.
Note that the expectation of the variable is also called {\it expected value}
(sometimes {\it expectation value}), {\it average} and {\it mean}. 
The square root of the variance is the {\it standard deviation} $\sigma$. 
\vspace{2ex}}
\begin{tabular}{lll}\hline
 & discrete variables & continuous variables \\
\hline
&& \nonumber \\
probability & $ P(X=x_i) = p(x_i)$ &
    $\intd P_{[x\le X \le x+\intd  x]}= p(x) \intd x $ \\
&& \nonumber \\
normalization$^\dag$ & $\sum_i p(x_i) = 1$
              &  $\int\!p(x)\intd x = 1$ \\
&& \nonumber \\
expectation of $f(X)$&  $\mbox{E}[f(X)] = \sum_i f(x_i) \, p(x_i) $
         &  $\mbox{E}[f(X)] = \int\!f(x) \, p(x)\intd x $ \\
&& \nonumber \\
expected value &  $\mbox{E}(X) = \sum_i x_i \, p(x_i) $
         &  $\mbox{E}(X) = \int\!x \, p(x)\intd x $ \\
&& \nonumber \\
moment of order $r$&  $\mbox{M}_r(X) = \sum_i x_i^r \, p(x_i) $
         &  $\mbox{M}_r(X) = \int\!x^r \, p(x)\intd x $ \\
&& \nonumber \\
variance & $\sigma^2 = \sum_i [x_i-\mbox{E}(X)]^2 \, p(x_i) $
         & $\sigma^2  = \int
            [x-\mbox{E}(X)]^2 \, p(x)\intd x $ \\
&& \nonumber \\
product rule   & $p(x_i,y_j) = p(x_i\,|\,y_j)\,p(y_j)$
               & $p(x,y) = p(x\,|\,y)\,p(y)$\\
&& \nonumber \\
independence & $p(x_i,y_j) = p(x_i)\,p(y_j)$
               & $p(x,y) = p(x)\,p(y)$\\
&& \nonumber \\
marginalization & $ \sum_jp(x_i,y_j) = p(x_i) $
                & $\int\!p(x,y)\intd y = p(x)$ \\
&& \nonumber \\
decomposition & $p(x_i) = \sum_jp(x_i\,|y_j)\,p(y_j)  $
                & $p(x) = \int\!p(x\,|\,y)\,
                  p(y)\intd y$ \\
&& \nonumber \\
Bayes' theorem & $p(x_j\,|\,y_i) =
                  \frac{\textstyle p(y_i\,|\,x_j)\,p(x_j)}
                      {\textstyle \sum_jp(y_i\,|\,x_j)\,p(x_j)}$  &
                 $p(x\,|\,y) =
                  \frac{\textstyle p(y\,|\,x)\,p(x)}
                      {\textstyle \int p(y\,|\,x)\,p(x)\intd x}$ \\
&& \nonumber \\
likelihood & ${\cal L}(x_j\,;\,y_i) = p(y_i\,|\,x_j) $ &
             ${\cal L}(x\,;\,y) = p(y\,|\,x) $ \\
&& \nonumber \\
\multicolumn{3}{l}
{$^\dag$\footnotesize A function $p(x)$ such that 
$\sum_i p(x_i) = \infty$, or
$\int\!p(x)\intd x = \infty$,  
is called {\it improper}. Improper} \\
\multicolumn{3}{l}{\hspace{2.0mm}\footnotesize functions are often used 
to describe {\it relative beliefs} about the possible 
values of a variable.} \\
\hline

\end{tabular}
\label{tab:fx}
\end{table}
\end{center}

\section{Bayesian inference for simple problems}
\label{sect:simplemodels}

We introduce the basic concepts of Bayesian inference 
by considering some
simple problems. 
The aim is to illustrate some of the notions that 
form the foundation of Bayesian reasoning. 

\subsection{Background information}
\label{sect:background}

As we think about drawing conclusions about the physical world, 
we come to realize that everything we do is based on what we 
know about the world.  Conclusions about hypotheses will
 be based on our general background knowledge.
To emphasize the dependence of probability on the state of
background information, which we designate as $I$, we will make 
it explicit by writing
$P(E\,|\,I)$, rather than simply $P(E)$. 
(Note that, in general, $P(A \,|\, I_1) \neq P(A \,|\, I_2) $, 
if $I_1$ and $I_2$ are different states of information.) 
For example, Eq.~(\ref{eq:joint}) should be more precisely
written as
\begin{equation}
 P(A\cap B\,|\,I) = P(A\,|\,B\cap I) \, 
P(B\,|\,I) = P(B\,|\,A\cap I) \,  P(A\,|\,I)\,,
\label{eq:joint_a} 
\end{equation}
or alternatively as 
\begin{equation}
 P(A, B\,|\,I) = P(A\,|\,B, I) \,  P(B\,|\,I) =
 P(B\,|\,A, I) \,  P(A\,|\,I) \,.
\label{eq:joint_b}
\end{equation}
We have explicitly included $I$ as part of the conditional to
remember that any probability relation is valid only under the same
state of background information.

\subsection{Bayes' theorem}\label{sec:BayesTheorem}
Formally, Bayes' theorem follows from the symmetry of $P(A,B)$ expressed by Eq.~(\ref{eq:joint_b}).
In terms of $E_i$ and $H_j$ belonging to two different
complete classes, Eq.~(\ref{eq:joint_b}) yields
\begin{equation}
\frac{P(H_j\,|\,E_i, I)}{P(H_j\,|\,I)} = \frac{P(E_i\,|\,H_j, I)}{P(E_i\,|\,I)}
\label{eq:bayes1} 
\end{equation}
This equation says that the {\it new} condition $E_i$ alters
our belief in $H_j$ by the same updating factor by which
the condition $H_j$ alters our belief about $E_i$.  
Rearrangement yields {\it Bayes' theorem}
\begin{equation}
P(H_j\,|\,E_i, I) = \frac{P(E_i\,|\,H_j, I) \,  P(H_j\,|\,I)}{P(E_i\,|\,I)}\,.
  \label{eq:bayes2}
\end{equation}
We have obtained a logical rule to update our beliefs on the basis of new conditions.
Note that, though Bayes' theorem is a direct consequence of the basic
rules of axiomatic probability theory, its updating power can  only  be fully exploited
if we can treat on the same basis expressions
concerning hypotheses and observations, causes and effects, models and data.

In most practical cases, the evaluation of $P(E_i\,|\,I)$ can be
quite difficult, while determining the conditional probability
$P(E_i\,|\,H_j,I)$ might be easier.  For example, think of $E_i$ as the probability
of observing a particular event topology in a particle physics
experiment, compared with the probability of
the same thing {\it given} a value of the hypothesized particle mass ($H_j$), a given
detector, background conditions, etc. Therefore, it is convenient to rewrite
$P(E_i\,|\,I)$ in Eq.~(\ref{eq:bayes2})
in terms of the quantities in the numerator,
using Eq.~(\ref{eq:decomp7}), to obtain
\begin{eqnarray}
P(H_j\,|\,E_i, I) &=& \frac{P(E_i\,|\,H_j, I) \,  P(H_j\,|\,I)}
                            {\sum_j P(E_i\,|\,H_j,I) \,  P(H_j\,|\,I)} \,,
  \label{eq:bayes3}
\end{eqnarray}
which is the better-known form of Bayes' theorem. Written this way, 
it becomes evident that the denominator of the r.h.s.\ 
of Eq.~(\ref{eq:bayes3}) 
is just a normalization factor and we can focus
on just the numerator:
\begin{eqnarray}
P(H_j\,|\,E_i,I) &\propto& P(E_i\,|\,H_j,I) \,  P(H_j \given I) \, .
  \label{eq:bayes5}
\end{eqnarray}
In words 
\begin{eqnarray}
\mbox{posterior} &\propto& \mbox{likelihood}\times \mbox{prior} \, ,
  \label{eq:bayes6}
\end{eqnarray}
where the {\it posterior} (or {\it final} state) stands for the probability
 of $H_j$, based on the
new observation $E_i$, relative to the {\it prior} (or {\it initial})  probability. 
(Prior probabilities are often indicated with $P_0$.)
The conditional probability $P(E_i\,|\,H_j)$ is called the
{\it likelihood}.  It is literally the probability of the observation 
$E_i$ given the specific hypothesis $H_j$. The term likelihood can 
lead to some confusion, because it is often misunderstood to mean 
``the likelihood that $E_i$ comes from $H_j$.'' 
However, 
 this name implies to consider $P(E_i\,|\,H_j)$ a mathematical function of 
$H_j$ for a fixed $E_i$ and in that framework it is usually written as 
${\cal L}(H_j; E_i)$ to emphasize the functionality.
We caution the reader that one sometimes even finds 
the notation 
${\cal L}(E_i\,|\,H_j)$ to indicate exactly $P(E_i\,|\,H_j)$. 

\subsection{Inference for simple hypotheses}

 Making use of formulae (\ref{eq:bayes3}) or (\ref{eq:bayes5}), we can easily
solve many classical problems involving inference when many hypotheses
can produce the same single effect.
Consider the case of interpreting the results of a test for the 
HIV virus applied to a {\it randomly
chosen} European. Clinical tests are very seldom perfect. 
Suppose that the test accurately detects infection, 
but has a false-positive rate of 0.2\%:
$$P(\mbox{Positive}\,|\,\mbox{Infected}) = 1 \, , \quad \mbox{and} \quad 
P(\mbox{Positive}\,|\,\mbox{$\overline{\mbox{Infected}}$}) = 0.2\%\,.$$
If the test is positive, can we conclude that the particular person 
is infected with a probability of 99.8\% 
 because the test has only a 0.2\% chance of mistake? Certainly not! This kind of mistake is often made by those who are not used to Bayesian reasoning, including scientists who make inferences in their own field of
expertise. The correct answer depends on 
what we else know about the person tested, that is, the background information.
Thus, we have to consider the incidence of the HIV virus in Europe, 
and possibly, information about the lifestyle of the individual.
For details, see (D'Agostini 1999c).

To better understand  the updating mechanism, let us take the ratio of 
Eq.~(\ref{eq:bayes3}) for two hypotheses $H_j$
and $H_k$
\begin{eqnarray}
\frac{P(H_j\,|\,E_i,I)}{P(H_k\,|\,E_i,I)} &=&
\frac{P(E_i\,|\,H_j,I)}{P(E_i\,|\,H_k,I)}
 \frac{P(H_j\,|\,I)}{P(H_k\,|\,I)}\, , \label{eq:bayes_factor}
\end{eqnarray}
where the sums in the denominators of Eq.~(\ref{eq:bayes3})  cancel. 
It is convenient to interpret the ratio of probabilities, 
given the same condition,
as {\it betting odds}. This is best done formally in 
the de Finetti approach,
but the basic idea is what everyone is used to:
the amount of money that one is willing
to bet on an event is proportional to the degree to which 
one expects that event will happen.
Equation~(\ref{eq:bayes_factor}) tells us that, when new 
information is available, the initial odds are updated
by the ratio of the likelihoods $P(E_i\,|\,H_j,I)/P(E_i\,|\,H_k,I)$, 
which is known as
the {\it Bayes factor}.

In the case of the HIV test, the initial odds for an 
arbitrarily chosen European to be infected $P(H_j\,|\,I)/P(H_k\,|\,I)$ are
so small that we need a very high Bayes' factor to
be reasonably certain that, when the test is positive, the person is
really infected.
With the numbers used in this example, the Bayes factor 
is $500 = 1/0.002$. For example, 
if we take for the prior
$P_0(\mbox{Infected})/P_0(\mbox{$\overline{\mbox{Infected}}$}) = 1/1000$,
the Bayes' factor changes these odds to $500/1000=1/2$, or equivalently, the
probability that the person is infected would be $1/3$, 
quite different from the $99.8\%$ answer usually prompted 
by those who have a standard statistical education. 
This example can be translated straightforwardly to physical problems, 
like particle identification in the analysis of a Cherenkov detector
data, as done, e.g. in (D'Agostini 1999c).

\section{Inferring numerical values of physics quantities --- General ideas 
and basic examples}
\label{sect:simple}
In physics we are concerned about models ('theories') and 
the numerical values of physical quantities related to them. 
Models and the value of quantities are, generally speaking,
the hypothesis we want to infer, given the observations.
In the previous section we have learned how to 
deal with simple hypotheses, `simple' in the sense
that they do not depend on internal parameters.  

On the other hand, in many applications we have strong beliefs about what
model to use to interpret the measurements.  Thus, we focus our
attention on the model parameters, which we consider as 
uncertain variables 
that we want to infer. The method which deals with these
applications is usually referred as {\it parametric inference},
and it will be shown with examples in this section.  
In our models, the value of the 
relevant physical quantities are usually described 
in terms of a continuous uncertain variable.  
Bayes' theorem, properly extended to uncertain quantities
(see Tab.\ref{tab:fx}), plays a central role in this inference process. 

A more complicate case is when we are also uncertain
about the model (and each possible model 
has its own set of parameter, usually associated 
with different physics quantities). We shall analyse this problem 
in Sect.~\ref{sect:modelcomp}.

\subsection{Bayesian inference on uncertain 
variables and posterior characterization}\label{sec:infvar}
We start here with a few  
one-dimensional problems involving
simple models that often occur in data analysis. 
These examples will be used to illustrate some of the 
most important Bayesian concepts. 
Let us first introduce briefly the structure of the Bayes' theorem
in the form convenient to our purpose, as a straightforward 
extension of what was seen in Sect.~\ref{sec:BayesTheorem}.
\begin{eqnarray}
p(\theta\,|\,d,I) &=& \frac{p(d\,|\,\theta,I)\,p(\theta\given I)}
                         {\int p(d\,|\,\theta,I)\,p(\theta\given I)\intd \theta}\,.
\label{eq:bayestheta}
\end{eqnarray}
 $\theta$ is the generic name of the parameter (used hereafter, unless 
the models have traditional symbols for their parameters) and $d$ is the
data point. $p(\theta\given I)$ is the prior, $p(\theta\,|\,d,I)$ the posterior
and $p(d\,|\,\theta,I)$ the likelihood. 
Also in this case the 
likelihood is is often written as
 ${\cal L}(\theta; d) = p(d\given \theta,I)$, and the same 
words of caution expressed in  Sect.~\ref{sec:BayesTheorem} apply here too.
Note, moreover, that, while $p(d\given \theta,I)$ is a properly
normalized pdf,  ${\cal L}(\theta; d)$ has not a pdf meaning in the
variable $\theta$. Hence,  
the integral of ${\cal L}( \theta; d)$ 
over $\theta$ is only accidentally equal to unity. 
The denominator in the r.h.s. of Eq.~(\ref{eq:bayestheta}) is called the {\it evidence} and, 
while in the parametric inference discussed here is just a trivial normalization factor,
its value becomes important for model comparison (see Sect. \ref{sect:modelcomp}). 

Posterior probability distributions
provide the full description of our state of knowledge about the
value of the quantity. In fact, they allow to calculate all
{\it probability intervals} of interest. Such intervals are 
also called {\it credible intervals} (at a specified level of probability,
for example 95\%)
or {\it confidence intervals} (at a specified level of 'confidence', 
i.e. of probability). 
However, the latter expression could be confused with 
frequentistic 'confidence intervals', that are not probabilistic
statements about uncertain variables (D'Agostini 1999c). 

It is often desirable to characterize the
distribution in terms of a few numbers. For example, 
mean value (arithmetic {\it average}) of the posterior,
or its most probable value (the {\it mode}) 
of the posterior, also known as the {\it maximum a 
posteriori (MAP) estimate}.   
The spread of the distribution is often
described in terms of its {\it standard deviation} 
(square root of the {\it variance}). 
It is useful to associate the terms mean value and standard deviation with
the more inferential terms {\it expected value},
or simply {\it expectation} (value), 
indicated by $\mbox{E}(\,)$,  and 
{\it standard uncertainty} (ISO 1993), indicated by $\sigma(\,)$, where the
argument is the uncertain variable of interest. 
This will be our standard way of reporting the result of
inference in a quantitative way, though, we emphasize 
that the full answer is given by the posterior distribution,
and reporting only these summaries 
in case of the complex distributions 
(e.g. multimodal and/or asymmetrical pdf's) 
can be misleading, because people tend to think of a 
Gaussian model if no further information is provided. 

\subsection{Gaussian model}
\label{sect:gaussian}
Let us start with a classical example in which the response signal $d$ 
from a detector is described by a Gaussian error function 
around the {\it true value} $\mu$ with a standard deviation $\sigma$, 
which is assumed to be exactly known. 
This model is the best-known among physicists and, indeed, 
the Gaussian pdf is also known as {\it normal} because
it is often assumed that  
errors are  'normally' distributed according to this function.  
Applying Bayes' theorem for continuous variables
(see Tab.~\ref{tab:fx}), from the likelihood
\begin{eqnarray}
p(d\,|\,\mu,I) &=& \frac{1}{\sqrt{2\pi}\,\sigma}\exp\left[
-\frac{(d-\mu)^2}{2\,\sigma^2}\right] \label{eq:Gaussian}
\end{eqnarray}
we get for $\mu$
\begin{eqnarray}
p(\mu\,|\,d,I) &=& \frac{\displaystyle\frac{1}{\sqrt{2\pi}\,\sigma}\exp\left[
-\frac{\displaystyle(d-\mu)^2}{2\,\sigma^2}\right] \,p(\mu\,|\,I)}
{\displaystyle\int_{-\infty}^{+\infty} \!
\frac{1}{\sqrt{2\pi}\,\sigma}\exp\left[
-\frac{(d-\mu)^2}{2\,\sigma^2}\right] \,p(\mu\,|\,I)\intd \mu
} \, .
\label{eq:infGaussian}
\end{eqnarray}
Considering all values of $\mu$ equally likely over a very large
interval, we can model the prior $p(\mu\,|\,I)$ with
a constant, which simplifies in Eq.~(\ref{eq:infGaussian}),
yielding
\begin{eqnarray}
p(\mu\,|\,d,I) &=& \frac{1}{\sqrt{2\pi}\,\sigma}\exp\left[
-\frac{(\mu-d)^2}{2\,\sigma^2}\right].  \label{eq:Gaussian_mu}
\end{eqnarray}
 Expectation and standard deviation of the posterior 
distribution  are $\mbox{E}(\mu)=d$
and $\sigma(\mu) = \sigma$, respectively. 
This particular result
corresponds to what is often done intuitively in practice. But
one has to pay attention to the assumed conditions under which the result
is logically valid:\ Gaussian likelihood and uniform prior.
Moreover, we can speak about the probability of true values only
in the subjective sense. It is recognized that physicists, and scientists
in general, are highly confused about this point (D'Agostini 1999a).

A noteworthy case of a prior for which the naive inversion
gives paradoxical results is when the value of a quantity is constrained
to be in the `physical region,' for example $\mu \ge 0$,
while $d$ falls outside it (or it is at its edge).
 The simplest prior that cures the problem
is a step function 
$\theta(\mu)$, 
and the result is
equivalent to simply renormalizing the pdf in the physical region
(this result corresponds to a `prescription' sometimes used by
practitioners with a frequentist background when they encounter
this kind of problem).

Another interesting case is when the prior knowledge can be
modeled with a Gaussian function, for example, describing our
knowledge from a previous inference
\begin{eqnarray}
p(\mu\,|\,\mu_0,\sigma_0,I) &=& \frac{1}{\sqrt{2\pi}\,\sigma_0}\exp\left[
-\frac{(\mu-\mu_0)^2}{2\,\sigma_0^2}\right] \,. \label{eq:Gaussian_prior}
\end{eqnarray}
Inserting Eq.~(\ref{eq:Gaussian_prior}) into
Eq.~(\ref{eq:infGaussian}), we get
\begin{eqnarray}
p(\mu\,|\,d,\mu_0,\sigma_0,I) &=& \frac{1}{\sqrt{2\pi}\,\sigma_1}\exp\left[
-\frac{(\mu-\mu_1)^2}{2\,\sigma_1^2}\right] \,, \label{eq:Gaussian_prior_inf}
\end{eqnarray}
where
\begin{eqnarray}
\mu_1 =\mbox{E}(\mu) &=& \frac{d/\sigma^2 + \mu_0/\sigma_0^2}
              {1/\sigma^2+1/\sigma_0^2}  \label{eq:mu_average} \\
 &=& \frac{\sigma_0^2}{\sigma^2+\sigma_0^2}\, d + 
     \frac{\sigma^2}{\sigma^2+\sigma_0^2}\, \mu_0 = 
\frac{\sigma_1^2}{\sigma^2}\,d + \frac{\sigma_1^2}{\sigma^2_0}\,\mu_0  
 \label{eq:mu_average_1}
\\
\sigma^2_1=\mbox{Var}(\mu) &=& \left(\sigma_0^{-2}+\sigma^{-2}\right)^{-1} 
\label{eq:sigma_average}
\end{eqnarray}
We can  then see that the
case $p(\mu\,|\,I)=\mbox{constant}$ corresponds
to the limit of a Gaussian prior 
with very large $\sigma_0$ and finite $\mu_0$.
The formula for the expected value combining 
previous knowledge and present experimental information
has been written in several ways in Eq.\,(\ref{eq:mu_average_1}). 

Another enlighting way of writing  Eq.\,(\ref{eq:mu_average}) is
considering $\mu_0$ and $\mu_1$ 
the estimates of $\mu$ at times $t_0$ and $t_1$, respectively
 before and after the observation $d$ happened at time $t_1$.
Indicating the {\it estimates} at different times by $\hat \mu(t)$, 
we can rewrite  Eq.\,(\ref{eq:mu_average}) as
\begin{eqnarray}
 \hat \mu(t_1) &=&  \frac{\sigma_\mu^2(t_0)}{\sigma_d^2(t_1)+\sigma_\mu^2(t_0)}\, d(t_1) + 
     \frac{\sigma_d^2(t_1)}{\sigma_d^2(t_1)+\sigma_\mu^2(t_0)}\, \hat\mu(t_0) 
  \nonumber \\
  &=&  \hat\mu(t_0) + \frac{\sigma_\mu^2(t_0)}{\sigma_d^2(t_1)+\sigma_\mu^2(t_0)}
                      \, [d(t_1) -  \hat\mu(t_0)] \\
  &=&  \hat\mu(t_0) + K(t_1) \, [d(t_1) -  \hat\mu(t_0)]
  \label{eq:Kalman1} \\
\sigma_\mu^2(t_1) &=& \sigma_\mu^2(t_0) -  K(t_1) \, \sigma_\mu^2(t_0) \,,
 \label{eq:Kalman2} 
\end{eqnarray} 
where 
\begin{eqnarray}
 K(t_1) &=& \frac{\sigma_\mu^2(t_0)}{\sigma_d^2(t_1)+\sigma_\mu^2(t_0)}\,.
\end{eqnarray} 
Indeed, we have given  Eq.\,(\ref{eq:mu_average}) the structure of a 
{\it Kalman filter} (Kalman 1960). The new observation `corrects' the
estimate by a quantity given by the {\it innovation} (or {\it residual}) 
$ [d(t_1) -  \hat\mu(t_0)]$ times the {\it blending factor} (or {\it gain}) 
$K(t_1)$. For an introduction about Kalman filter and its probabilistic
origin, see (Maybeck 1979 and Welch and Bishop 2002).      

As Eqs.~(\ref{eq:mu_average_1})--(\ref{eq:Kalman2}) show, a new experimental 
information reduces the uncertainty. But this is true as long 
the previous information and the observation are somewhat consistent.
If we are, for several reasons, sceptical about the model which yields
the combination rule (\ref{eq:mu_average_1})--(\ref{eq:sigma_average}),
we need to remodel the problem and introduce possible
systematic errors or underestimations of the quoted standard deviations, 
as done e.g. in (Press 1997, Dose and von der Linden 1999, 
D'Agostini 1999b, Fr\"ohner 2000).

\subsection{Binomial model}
\label{sect:counting}
\label{sect:binomial}    
In a large class of experiments, the observations consist of counts, 
that is, a number of things (events, occurrences, etc.).
In many processes of physics interests the 
resulting number of counts is described probabilistically by a binomial
or a Poisson model. 
For example, we want to draw an 
inference about the efficiency of a detector, a branching ratio 
in a particle decay or a rate from a measured number of counts
in a given interval of time. 

The binomial distribution describes
 the probability of randomly obtaining $n$ events (`successes') 
in $N$ independent trials, in each of which we assume the same probability 
 $\theta$ that the event will happen. The probability function is
\begin{equation}
p(n\,|\,\theta,N) = \left(\!\!\begin{array}{c}N \\ n\end{array}\!\!\right)
                     \theta^n(1-\theta)^{N-n}\,,
\label{eq:binomial}
\end{equation}
where the leading factor is the well-known binomial coefficient, 
namely ${N!}/{n!(N-n)!}$. 
We wish to infer $\theta$ from an observed number of counts $n$ in $N$ trials.  
Incidentally, that was the
``problem in the doctrine of chances'' originally treated by Bayes
(1763), reproduced e.g. in (Press 1992). Assuming a uniform prior for $\theta$, 
by Bayes' theorem the posterior distribution for $\theta$ is 
proportional to the likelihood, given by Eq.~(\ref{eq:binomial}):
\begin{eqnarray}
p(\theta \,|\,n,N,I) & = &
   \frac{ \theta^n\,(1-\theta)^{N-n}}
 {\int_0^1 \theta^n\,(1-\theta)^{N-n}\intd \theta} \\
 & = & \frac{(N+1)!}{n!\,(N-n)!}\,\theta^n\,(1-\theta)^{N-n}\,.
 \label{eq:inv_binom}
\end{eqnarray}
\begin{figure}
\begin{center}
\includegraphics[width=0.7\linewidth]{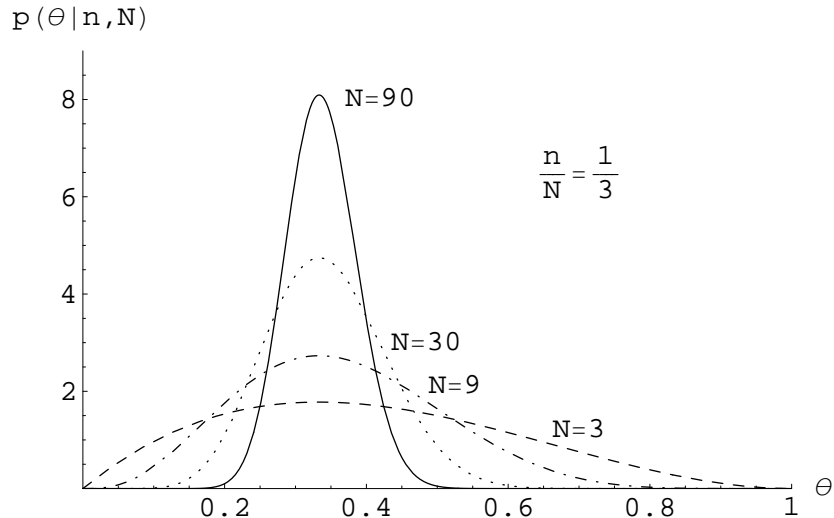}
\end{center}
\caption{Posterior probability density function of the binomial parameter
$\theta$, having observed $n$ successes in $N$ trials.}
\label{fig:beta}
\end{figure}
Some examples of this distribution for various values of $n$ and $N$ 
are shown in Fig.~\ref{fig:beta}.
Expectation, variance, and mode of this distribution are:
\begin{eqnarray}
\mbox{E}(\theta) &=& \frac{n+1}{N+2}
\label{eq:infbinom1}\\
\sigma^2(\theta) &=& 
        \frac{(n+1)(N-n+1)}{(N+3)(N+2)^2}
       = \frac{\mbox{E}(\theta)\,\left(1 - \mbox{E}(\theta)\right)}{N+3} \\
\theta_{\mbox{\footnotesize m}} &=& \frac{n}{N}\,, 
\label{eq:infbinom2}
\end{eqnarray}
where the mode has been indicated with $\theta_{\mbox{\footnotesize m}}$.
Equation~(\ref{eq:infbinom1}) is known as the Laplace formula. 
For large values of $N$ and $0 \ll n \ll N$
the expectation of $\theta$ tends to $\theta_{\mbox{\footnotesize m}}$, 
 and $p(\theta)$ becomes approximately Gaussian. 
This result is nothing but a reflection
of the well-known asymptotic Gaussian behavior of $p(n\given \theta,N)$.
For large $N$ the uncertainty about
$\theta$ goes like $1/\sqrt{N}$. Asymptotically, we are practically
certain that $\theta$ is equal to the relative frequency of that class
of events observed in the past. This is how the frequency based 
evaluation of probability is promptly recovered in the Bayesian approach,
under well defined assumptions.

\begin{figure}
\begin{center}
\includegraphics[width=0.7\linewidth]{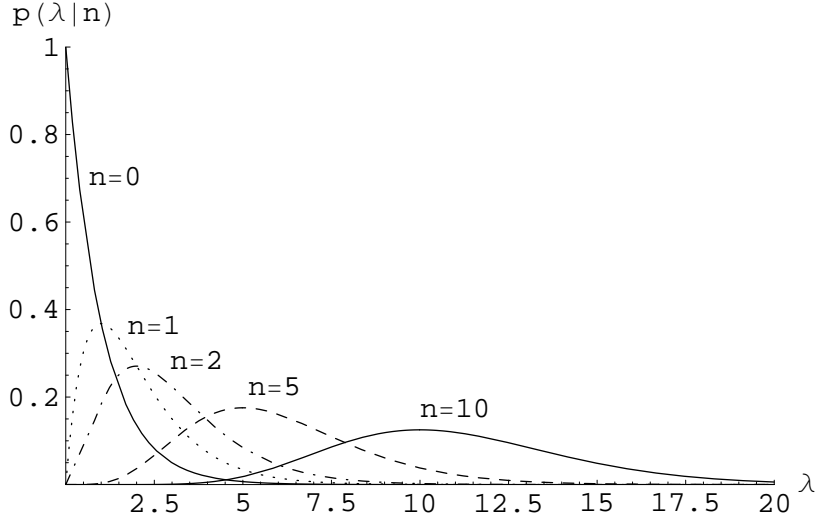}
\end{center}
\caption{The posterior distribution for the Poisson parameter $\lambda$, 
when $n$ counts are observed in an experiment.}
\label{fig:invpois}
\end{figure}
\subsection{Poisson model}
\label{sect:poisson}
The Poisson distribution gives the probability of observing $n$ 
counts in a fixed time interval, 
when the expectation of the number of counts to be observed is $\lambda$:
\begin{eqnarray}
p(n\,|\,\lambda) &=& \frac{\lambda^n e^{-\lambda}}{n!}\,.
\end{eqnarray}
The inverse problem is to infer $\lambda$ from $n$ counts observed. 
(Note that what physically matters is the rate $r=\lambda/\Delta T$,
where $\Delta T$ is the observation time.)  
Applying Bayes' theorem and using a uniform prior 
$p(\lambda\given I)$ for $\lambda$, we get
\begin{equation}
p(\lambda\,|\,n,I) =  \frac{\displaystyle\frac{\lambda^n \, e^{-\lambda}}{n!}}
{\displaystyle\int_0^\infty{\frac{\lambda^n\,e^{-\lambda}}{n!}
 \,\rm {d}\lambda}}
 = \frac{\lambda^n\, e^{-\lambda}}{n!} \,.
 \label{eq:inv_poiss1}
\end{equation}
As for the Gaussian model, the same mathematical
expression holds for the likelihood, but with interchanged role of variable
and parameter. Expectation and variance of $\lambda$ are both
equal to $n+1$, while the most probable value is $\lambda_m =n$. 
For large $n$, the extra `$+1$' (due to the asymmetry of
the prior with respect to $\lambda=0$) can be ignored and we have
$\mbox{E}(\lambda)=\sigma^2(\lambda)\approx n$ and, once again, 
the uncertainty about $\lambda$ follows a Gaussian model.
The relative uncertainty on $\lambda$ decreases as $1/\sqrt{n}$. 

When the observed value of $n$ is zero, Eq.~(\ref{eq:inv_poiss1})
yields $p(\lambda\,|\,n=0)=e^{-\lambda}$, giving a maximum of
belief at zero, but an exponential tail toward large values of
$\lambda$. Expected value and standard deviation of $\lambda$ 
are both equal to 1. 
The  95\% probabilistic upper bound of $\lambda$ is at
$\lambda_{95\% UB}=3$, as it can be easily calculated solving
the equation 
$\int_0^{\lambda_{95\% UB}}\!\!p(\lambda\,|\,n=0)\intd \lambda=0.95$. 
Note that also this result depends on the
choice of prior, though Astone and D'Agostini (1999) have 
shown that the upper bound is insensitive to
the exact form of the prior, if the prior models somehow what
they call ``positive attitude of rational scientists'' 
(the prior has not to be in contradiction with what one could actually observe, 
given the detector sensitivity). In particular, they show that a uniform prior
is a good practical choice to model this attitude. 
On the other hand, talking about `objective' probabilistic upper/lower
limits makes no sense, as discussed in detail and with examples in
the cited paper:\ one can at most 
speak about conventionally defined non-probabilistic {\it sensitivity bounds}, 
which separate the measurement region from that in which 
experimental sensitivity is lost (Astone and D'Agostini 1999, D'Agostini 2000, 
Astone \etal 2002).

\subsection{Inference from a data set and sequential use of Bayes formula}
In the elementary examples shown above, the inference has been done
from a single {\it data point} $d$. If we have a set of observations
(data), indicated by $\mvec d$, we just need to insert in the Bayes formula
the likelihood $p(\mvec d\given\theta ,I)$, where this expression indicates
a multi-dimensional joint pdf. 

Note that we could think of inferring $\theta$ on the basis of
each newly observed datum $d_i$. After the one observation:
\begin{equation}
p(\theta  \given d_1,I) \propto 
   p(d_1 \given \theta,I) \, p(\theta \,|\, I)  \label{eq:iter1}
\end{equation}
and after the second:
\begin{eqnarray}
p(\theta  \given d_1,d_2,I) &\propto &
   p(d_2 \given \theta,d_1,I) \, p(\theta  \given d_1,I) \\
  &\propto &  p(d_2 \given \theta,d_1,I) \, p(d_1 \given \theta,I)
   \, p(\theta \,|\, I)\,. \label{eq:iter2}
\end{eqnarray}
We have written Eq.~(\ref{eq:iter2}) in a way that the 
dependence between observables can be accommodated. From the product rule in
Tab.~\ref{tab:fx}, we can rewrite Eq.~(\ref{eq:iter2}) as
\begin{eqnarray}
p(\theta  \given d_1,d_2,I) &\propto &
p(d_1,d_2 \given \theta,I) \,p(\theta \,|\, I)\,.
\end{eqnarray}
Comparing this equation with (\ref{eq:iter2}) 
we see that the sequential inference gives exactly the same result of a
single inference that properly takes into account
all available information. This is an important result
of the Bayesian approach. 

The extension to many variables is straightforward, 
obtaining
\begin{eqnarray}
p(\mvec\theta\given \mvec d, I) &\propto & 
p(\mvec d\given \mvec\theta,I) \, p(\mvec\theta\given I) \,.
\end{eqnarray}
Furthermore, when the $d_i$ are independent, we get for the likelihood
\begin{eqnarray}
p(\mvec d \given \mvec \theta,I) &=& \prod_i p(d_i \given \mvec \theta,I) 
\label{eq:likproduct} \\
{\cal L}(\mvec \theta; \mvec d) &=& \prod_i {\cal L}(\mvec \theta;  d_i) \, ,
\end{eqnarray}
that is, the combined likelihood is given by the product 
of the individual likelihoods.

\subsection{Multidimensional case --- Inferring $\mu$ and $\sigma$ of a Gaussian}
\label{sec:MultidimensionalCase}

So far we have only inferred one parameter of a model.
The extension to many parameters
is straightforward. Calling $\mvec \theta$ the set of parameters
and $\mvec d$ the data, Bayes' theorem becomes
\begin{eqnarray}
p(\mvec \theta  \given \mvec d,I) &=&
\frac{p(\mvec d  \given \mvec \theta,I) \, p(\mvec \theta \,|\, I)}
{\int p(\mvec d  \given \mvec \theta,I) \, p(\mvec \theta \,|\, I) 
\intd \mvec \theta}
\,. \label{eq:multi}
\end{eqnarray}
Equation~(\ref{eq:multi}) gives the posterior for the full 
parameter vector $\mvec \theta$. 
{\it Marginalization} (see Tab.~\ref{tab:fx})
allows one to
calculate the probability distribution for a single parameter, 
for example, $p(\theta_i \given \mvec d , I)$, 
by integrating over the remaining
parameters. The marginal distribution $p(\theta_i \given \mvec d, I)$ 
is then the complete result of the Bayesian
inference on the parameter $\theta_i$. Though the characterization
of the marginal is done in the usual way described in 
Sect.~\ref{sec:infvar}, there is often the interest to 
summarize some characters of the multi-dimensional 
posterior that are unavoidably lost in the marginalization
(imagine marginalization as a kind of geometrical projection).
Useful quantities are the covariances between parameters 
$\theta_i$ and $\theta_j$, defined as
\begin{eqnarray}
\mbox{Cov}(\theta_i,\theta_j) &=& 
\mbox{E}[(\theta_i-\mbox{E}[\theta_i])\,(\theta_j-\mbox{E}[\theta_j])]\,.
\end{eqnarray}
As is well know, quantities which give a more intuitive idea 
of what is going on are the correlation coefficients, defined
as $\rho(\theta_i,\theta_j)=
\mbox{Cov}(\theta_i,\theta_j)/\sigma(\theta_i)\sigma(\theta_j)$.
Variances and covariances form the 
covariance matrix  
$\mathbf{V}(\mvec{\theta})$, with $V_{ii}=\mbox{Var}(\theta_i)$
and $V_{ij} = \mbox{Cov}(\theta_i,\theta_j)$.
We recall also that convenient formulae to calculate variances and covariances
are obtained from the expectation of the products $\theta_i\theta_j$,
together with the expectations of the parameters:
\begin{eqnarray}
V_{ij} &=& \mbox{E}(\theta_i\theta_j) - \mbox{E}(\theta_i)\,\mbox{E}(\theta_j)
\label{eq:calc_cov}
\end{eqnarray}
As a first example of a multidimensional distribution from a
data set, we can think, again, at the inference of the parameter 
$\mu$  of a Gaussian distribution, but in the case that also
$\sigma$ is unknown and needs to be determined by the data. 
From Eqs. (\ref{eq:multi}), (\ref{eq:likproduct}) and 
(\ref{eq:Gaussian}), with $\theta_1=\mu$ and $\theta_2=\sigma$
and neglecting overall normalization, we obtain
\begin{eqnarray}
p(\mu,\sigma\given \mvec d, I) & \propto & 
\sigma^{-n}\,\exp\left[-\frac{\sum_{i=1}^n(d_i-\mu)^2}{2\,\sigma^2}\right]
\,p(\mu,\sigma\given I) \label{eq:inf_mu_sigma} \\
p(\mu\given \mvec d, I) & = & \int p(\mu,\sigma\given \mvec d, I)
\intd \sigma  \label{eq:inf_mu_sigma1} \\
p(\sigma\given \mvec d, I) & = & \int p(\mu,\sigma\given \mvec d, I)
\intd \mu\,.  \label{eq:inf_mu_sigma2}
\end{eqnarray}
The closed form of Eqs.~(\ref{eq:inf_mu_sigma1}) and \ref{eq:inf_mu_sigma2}) 
depends on the prior and, perhaps, for 
the most realistic choice of $p(\mu,\sigma\given I)$, such a 
compact solution does not exists. But this is not an essential issue, 
given the present computational power. 
(For example, the shape of  $p(\mu,\sigma\given I)$ can be easily inspected
by a modern graphical tool.)  
We want to stress here
the conceptual simplicity of the Bayesian solution to the problem.
[In the case the data set contains some more than a dozen of
observations, a flat $p(\mu,\sigma\given I)$, with the constraint
$\sigma > 0$, can be considered a good practical choice.]

\subsection{Predictive distributions}
\label{sec:predictive}
A related problem is to `infer' what an experiment will observe
given our best knowledge of the underlying theory and its parameters.
Infer is within quote marks because the term is usually used for 
model and parameters, rather than for observations. In this
case people prefer to speak about  {\it prediction} (or {\it prevision}). 
But we recall that in the Bayesian reasoning there is conceptual 
symmetry between the  uncertain quantities which enter the
problem. Probability density functions  describing not yet observed
event are referred to as {\it predictive distributions}. 
There is a conceptual difference with the likelihood, which
also gives a probability of observation, but under different hypotheses,
as the following example clarifies. 

Given $\mu$ and $\sigma$, and assuming a Gaussian model,
our uncertainty about a `future' $d_f$ is described by the Gaussian
pdf Eq.~(\ref{eq:Gaussian}) with $d=d_f$. But this holds only 
under that particular hypothesis for $\mu$ and $\sigma$, 
while, in general, we are also uncertain about these values too.
Applying the decomposition formula (Tab.~\ref{tab:fx}) we get: 
\begin{eqnarray}
p(d_f\given I) &=& \int p(d_f\given \mu,\sigma,I)\,p(\mu,\sigma\given I)
\intd\mu\intd\sigma
\label{eq:ddGivenI}
\end{eqnarray}
Again, the integral might be technically difficult, but the solution
is conceptually simple. Note that, though the decomposition formula
is a general result of probability theory, it can be applied to
this problem only in the subjective approach. 

An analytically easy,  insightful case is that of experiments with 
well-known $\sigma$'s. Given a past observation $d_p$ 
and a vague prior, $p(\mu\given d_p,I)$ is Gaussian around $d_p$
with variance $\sigma^2_p$
[note that, with respect to $p(\mu,\sigma\given I)$ of Eq.(\ref{eq:ddGivenI}),  
it has been made explicit that $p(\mu)$ depend on  $d_p$].
$p(d_f\given \mu)$ is Gaussian around $\mu$
with variance $\sigma^2_f$. We get finally
\begin{eqnarray}
p(d_f\given d_p,I) &=& \int p(d_f\given \mu,I)\,p(\mu\given d_p,I)\intd \mu \\
                   &=& \frac{1}{\sqrt{2\pi}\,\sqrt{\sigma_p^2+\sigma_f^2}} 
  \,\exp \left[-\frac{(d_f-d_p)^2}{2\,(\sigma_p^2+\sigma_f^2)}\right] \,.                 
\end{eqnarray}

\subsection{Hierarchical modelling and hyperparameters}
\label{sect:hierarchical}
As we have seen in the previous section, 
it is often desirable to include in a probabilistic model one's 
uncertainty in various aspects of a pdf.  
This is a natural feature of the Bayesian methods, 
due to the uniform approach to deal with uncertainty
and from which powerful analysis tools are derived. 
This kind of this modelling is called 
{\it hierarchical} because the characteristics 
of one pdf are controlled by another pdf. 
All uncertain parameters from which the pdf
depends are called {\it hyperparameter}. 
An example of use of hyperparameter is described in 
 Sect.~\ref{sect:conjugate}  in which the 
prior to infer $\theta$ in a binomial model are shown to be controlled
by the parameters of a Beta distribution. 

As an example of practical importance, think of 
the combination of experimental results in the presence 
of {\it outliers}, i.e. of data points which are 
somehow in mutual disagreement. In this case the combination 
rule given by Eqs.~(\ref{eq:mu_average})--(\ref{eq:sigma_average}),
extended to many data points, produces unacceptable conclusions.
A way of solving the problem (Dose and von der Linden 1999, 
D'Agostini 1999b) is to model a scepticism about the quoted 
standard deviations of the experiments, introducing 
a pdf $f(r)$, where $r$ is a rescaling factor of the 
standard deviation. In this way the $\sigma$'s that enter
the r.h.s. of Eqs.~(\ref{eq:mu_average})--(\ref{eq:sigma_average})
are hyperparameters of the problem. 
An alternative approach, 
also based on hierarchical modelling, is shown in (Fr\"ohner 2000). 
For a more complete introduction to the subject see e.g.
(Gelman \etal 1995).  

\subsection{From Bayesian inference to maximum-likelihood 
and minimum chi-square model fitting}
\label{sect:maxlik}

Let us continue with the case in which we know so little about 
appropriate values of the parameters
that a uniform distribution is a practical choice for the prior. 
Equation~(\ref{eq:multi})
becomes 
\begin{eqnarray}
p(\mvec \theta  \given \mvec d,I) &\propto &
p(\mvec d  \given  \mvec \theta,I) \, p_0(\mvec \theta,I)
\propto p(\mvec d  \given \mvec \theta,I)
= {\cal L}(\mvec \theta;  \mvec d)
\,, \label{eq:multi2}
\end{eqnarray}
where, we recall, the likelihood ${\cal L}(\mvec \theta;  \mvec d)$
is seen as 
 a mathematical function of $\mvec \theta$, with parameters $\mvec d$.

The set of $\mvec \theta$ that is most likely is that which maximizes
${\cal L}(\mvec \theta;  \mvec d)$, a result known as the
{\it maximum likelihood principle}. Here it has been
obtained again as a special case of a more general
framework,  under
clearly stated hypotheses, without need to introduce new ad hoc rules. 
Note also that the inference does not depend
on multiplicative factors in the likelihood.
This is one of the ways to state the 
{\it likelihood principle}, ideally desired by frequentists,
but often violated. This `principle' always and naturally
holds in Bayesian statistics. 
It is important to remark that
the use of unnecessary principles is dangerous, because there
is a tendency to use them
uncritically. For example, formulae resulting from
maximum likelihood are often used also when
 non-uniform reasonable priors should be
taken into account, or when the shape of ${\cal L}(\mvec \theta;  \mvec d)$
is far from being multi-variate Gaussian. (This is
a kind of ancillary
default hypothesis that comes together with this principle,
and is the source of the often misused `$\Delta (-\ln {\cal L}) = 1/2$' rule
to determine probability intervals.)

The usual least squares formulae are easily
derived if we take the
well-known case of pairs $\{x_i,y_i\}$ 
(the generic $\mvec d$ stands for all data points)
whose true values are related by a deterministic function
$\mu_{y_i} = y(\mu_{x_i},\mvec \theta)$ and
with Gaussian errors only in the ordinates, i.e.
we consider $x_i \approx \mu_{x_i}$.
In the case of independence of the measurements, the 
likelihood-dominated result becomes,
\begin{eqnarray}
p(\mvec \theta  \given \mvec x,\mvec y,I)
&\propto& \prod_i \exp\left[
  -\frac{(y_i-y(x_i,\mvec \theta))^2}{2\,\sigma_{i}^2}\right]
\end{eqnarray}
or
\begin{eqnarray}
p(\mvec \theta  \given \mvec x,\mvec y,I)
&\propto& \exp\left[-\frac{1}{2}\chi^2\right] \,,
\label{eq:expchi2}
\end{eqnarray}
where 
\begin{eqnarray}
\chi^2(\mvec \theta)&=&\sum_i \frac{(y_i-y(x_i,\mvec \theta))^2}{\sigma_{i}^2}
 \label{eq:chi2}
\end{eqnarray}
is called `chi-square,' well known among physicists.  
Maximizing the likelihood is equivalent to
minimizing $\chi^2$, and the most probable value
of $\mvec \theta$ is easily obtained (i.e. the {\it mode}
indicated with $\mvec \theta_m$), analytically in easy cases,
or numerically for more complex ones.

As far as the uncertainty in  $\mvec \theta$ is concerned,
the widely-used evaluation of the covariance matrix $\mathbf{V}(\mvec \theta)$ 
(see Sect. \ref{sec:MultidimensionalCase}) 
from the Hessian,
\begin{eqnarray}
(V^{-1})_{ij} (\mvec \theta) &=& \left. \frac{1}{2}
\frac{\partial^2\chi^2}{\partial\theta_i\partial\theta_j}
\right|_{\mvec\theta=\mvec\theta_m}\,,
\label{eq:invV}
\end{eqnarray}
 is merely consequence
of an {\it assumed}
 multi-variate Gaussian distribution of  $\mvec \theta$, 
that is   a parabolic shape of  $\chi^2$ 
(note that the `$\Delta (-\ln {\cal L}) = 1/2$' rule,
and the from this rule resulting `$\Delta \chi^2 = 1$ rule,' 
has the same motivation).
In fact,  expanding  
$\chi^2(\mvec\theta)$ in series around its minimum, we have
\begin{eqnarray}
\chi^2(\mvec\theta) &\approx&  \chi^2(\mvec\theta_m) + 
              \frac{1}{2}\, \mvec{\Delta}\theta^T\, \mathbf{H}\,\mvec{\Delta}\theta
\label{eq:chi2_approx}
\end{eqnarray}
where $\mvec{\Delta}\theta$ stands for the the set of differences 
$\theta_i-\theta_{m_i}$ and $\mathbf{H}$ is the Hessian matrix, whose elements
are given by twice the r.s.h. of 
Eq.~(\ref{eq:invV}). Equation\ (\ref{eq:expchi2}) becomes then
\begin{eqnarray}
p(\mvec \theta  \given \mvec x,\mvec y,I)
&\approx& \propto \exp\left[-\frac{1}{2}  
                    \mvec{\Delta}\theta^T\, \mathbf{H}\,\mvec{\Delta}\theta \right] \,,
\label{eq:expchi2_approx}
\end{eqnarray}
which we recognize to be a multi-variate Gaussian distribution
if we identify $\mathbf{H}=\mathbf{V}^{-1}$. 
After normalization, we get finally
\begin{eqnarray}
p(\mvec \theta \given \mvec x,\mvec y,I) &\approx& 
(2 \pi)^{-n/2}\, (\det\mathbf{V})^{-1/2}\,
   \exp \left[-\frac{1}{2}\mvec{\Delta}\theta^T \,
   \mathbf{V}^{-1}\mvec{\Delta}\theta
   \right] \,,
\label{eq:multi-variate-Gaussian}
\end{eqnarray}
with $n$ equal to the dimension of $\mvec\theta$ and 
$\det\mathbf{V}$ indicating the  determinant of $\mathbf{V}$.
Holding this approximation, $\mbox{E}(\mvec\theta)$ 
is approximately equal to $\mvec\theta_m$.
Note that the result (\ref{eq:multi-variate-Gaussian})
is exact when $y(\mu_{x_i},\mvec\theta)$ depends
linearly on the various $\theta_i$. 

In routine applications, the hypotheses that lead to the
maximum likelihood and least squares formulae often hold.
But when these hypotheses are not justified, we need
to characterize the result by the multi-dimensional posterior distribution
$p(\mvec \theta)$, going back to the more general expression
Eq.~(\ref{eq:multi}).

The important conclusion from this section,
as was the case for the definitions of probability
 in Sect.~\ref{sect:rules}, is that
Bayesian methods often lead to well-known
conventional results, but without introducing them as
new ad hoc rules as the need arises. The analyst acquires 
then a heightened sense of awareness about the range 
of validity of the methods. One might as well use these
`recovered' methods within the Bayesian framework, with its
more natural interpretation
of the results.  Then one can speak about 
the uncertainty in the model parameters and quantify it with
probability values, which is the usual 
way in which physicists think.

\subsection{Gaussian approximation of the posterior distribution}
\label{GaussianPosterior}
The substance of the results seen in the previous section
holds also in the case in which the prior is not flat and, hence,
cannot be absorbed in the normalization constant of the posterior. 
In fact, in many practical cases the posterior exhibits an approximately
(multi-variate) Gaussian shape, even if the prior was not trivial. 
Having at hand an {\it un-normalized} posterior $\tilde p()$, i.e. 
\begin{eqnarray}
\tilde p(\mvec \theta \given \mvec d,I) &=& 
p(\mvec d  \given  \mvec \theta,I) \, p_0(\mvec \theta,I)\,,
\label{eq:unnormposterior}
\end{eqnarray}
we can take its {\it minus-log} function 
$\varphi(\mvec \theta ) = -\ln \tilde p(\mvec \theta)$.
If $\tilde p(\mvec \theta \given \mvec x,\mvec y,I)$ has 
approximately a Gaussian shape, it follows that 
\begin{eqnarray}
\varphi(\mvec \theta ) &\approx& 
\frac{1}{2}\,\mvec{\Delta}\theta^T \,
   \mathbf{V}^{-1}\mvec{\Delta}\theta + \mbox{\it constant}\,.
\end{eqnarray}
$\mathbf{V}$ can be evaluated as
\begin{eqnarray}
(V^{-1})_{ij} (\mvec \theta) &\approx& \left.
\frac{\partial^2\varphi}{\partial\theta_i\partial\theta_j}
\right|_{\mvec\theta=\mvec\theta_m}\,,
\label{eq:invVlogp}
\end{eqnarray}
where $\mvec\theta_m$ was obtained from the 
minimum of $\varphi(\mvec\theta)$.

\section{Uncertainties from systematic effects}
\label{sect:systematic}

The uncertainty described in the previous section are related
to the so-called {\it random}, or {\it statistical} errors. 
Other important sources are, generally speaking (see ISO 1993 for
details), related to uncertain values of {\it influence variables}
on which the observed values, or the data-analysis process, might
depend. In physics, we usually refer to these as systematic effects or errors.
They can be related to the parameters of the 
experiment, like a particle beam energy or the exposure time, 
or to environmental variables, like temperature and
pressure, calibration constants of the detector, and all other
parameters,  `constants'  (in the physical sense), and hypotheses
that enter the data analysis. The important thing is that
we are unsure about their precise value.
Let us indicate all the influence variables 
with the vector $\mvec h = \{h_1, h_2, \ldots, h_n\}$, and their 
joint pdf as $p(\mvec h \given I)$. 

The treatment of uncertainties due to systematic errors has
traditionally been lacking  a consistent theory, essentially
due the unsuitability to standard statistical methods of 
dealing with uncertainty in the most wide sense.
Bayesian reasoning becomes crucial to handle these sources
of uncertainty too, and even metrological organizations (ISO 1993)
had to recognized it. For example, the ISO {\it type B} uncertainty
is recommended to be {\it ``evaluated by scientific judgment based on all
the available information on the possible variability''} (ISO 1993)
of the influence quantities (see also D'Agostini 1999c).

\subsection{Reweighting of conditional inferences}
\label{sect:reweighting}

The values of the influence variables and their uncertainties 
contribute to our background knowledge $I$ about the experimental 
measurements. Using $I_0$ to represent our very general background 
knowledge, the posterior pdf
will then be $p(\mu \given \mvec d,\mvec h,I_0)$, where the dependence on
all possible values of $\mvec h$ has been made explicit. The
inference that takes into account the uncertain vector $\mvec h$ is
obtained using the rules of probability (see Tab.~\ref{tab:fx}) 
by integrating the joint probability over the uninteresting 
influence variables:
\begin{eqnarray}
p(\mu \given \mvec d, I_0) &=& 
 \int\! p(\mu,\mvec h \given \mvec d, I_0) \intd \mvec h \\
&=& \int\! p(\mu \given \mvec d,\mvec h,I_0)
\,p(\mvec h \given I_0)\intd \mvec h \,.
\label{eq:syst_gen}
\end{eqnarray}
As a simple, but important case, let us consider a single
influence variable given by an additive instrumental offset
 $z$, which is expected to
be zero because the instrument has been calibrated as well as feasible and 
the remaining uncertainty is $\sigma_z$.  Modelling our
uncertainty in $z$ as a Gaussian distribution with a standard
deviation $\sigma_z$, the posterior for $\mu$ is
\begin{eqnarray}
p(\mu \given d,I_0) &=&
\int_{-\infty}^{+\infty}\!p(\mu \given d,z,\sigma,I_0)
\,p(z\,|\sigma_z,I_0)\intd z \\
 &=& \int_{-\infty}^{+\infty}\!
\frac{1}{\sqrt{2\pi}\,\sigma}\exp\left[
-\frac{(\mu-(d-z))^2}{2\,\sigma^2}\right] \times \, \nonumber \\
& & \ \ \ \ \ \ \  \frac{1}{\sqrt{2\pi}\,\sigma_z}\exp\left[
-\frac{z^2}{2\,\sigma_z^2}\right] \intd z \\
 &=& \frac{1}{\sqrt{2\pi}\,\sqrt{\sigma^2+\sigma_z^2}}\exp\left[
-\frac{(\mu-d)^2}{2\,(\sigma^2+\sigma_z^2)}\right] \, .
\label{eq:syst_z}
\end{eqnarray}
The result is that the net variance is the sum of 
the variance in the measurement and the variance in the influence variable.

\subsection{Joint inference and marginalization of
nuisance parameters}
\label{sect:glob_inf}

A different approach, which produces identical results,
is to think of a joint
inference about both the quantities of interest
and the influence variables: 
\begin{eqnarray}
p(\mu,\mvec h \given \mvec d,I_0) &\propto& p(\mvec d \given  \mu,\mvec h,I_0) \,
   p_0(\mu,\mvec h \given I_0) \,.
\label{eq:globinf}
\end{eqnarray}
Then, marginalization is applied to the variables that we are not
interested in (the so called {\it nuisance parameters}), obtaining
\begin{eqnarray}
p(\mu|\,\mvec d,I_0) &=& 
\int\!p(\mu,\mvec h \given \mvec d,I_0)\intd \mvec h \\
&\propto& \int p(\mvec d \given  \mu,\mvec h,I_0) \,
   p_0(\mu,\mvec h \given I_0) \intd \mvec h \,.
\label{eq:globinf_marg}
\end{eqnarray}
Equation~(\ref{eq:globinf}) shows a peculiar feature of Bayesian inference,
namely the possibility making an inference about a number of
variables larger than the number of the observed data. Certainly,
there is no magic in it, and the resulting variables will be highly
correlated. Moreover, the prior cannot be improper in all
variables. 
But, by using {\it informative priors} in which experts feel
confident,
this feature allows one to tackle complex problems with
missing or corrupted parameters. In the end, making use of
marginalization, one can concentrate on the quantities of real interest.

The formulation of the problem in terms of
Eqs.~(\ref{eq:globinf}) and (\ref{eq:globinf_marg}) allows one to
solve problems in which the influence variables might depend
on the true value $\mu$, because $ p_0(\mu,\mvec h \given I_0)$
can model dependences between $\mu$ and $\mvec h$. In most
applications,  $\mvec h$ does not depend on $\mu$, and the
prior factors into the product of
 $p_0(\mu \given I_0)$ and $p_0(\mvec h \given I_0)$. When this happens,
we recover exactly the same results as obtained using
the reweighting of conditional inferences approach
described just above.

\subsection{Correlation in results caused by systematic errors}

We can easily extend Eqs.~(\ref{eq:syst_gen}),
(\ref{eq:globinf}), and (\ref{eq:globinf_marg})
to a joint inference of several variables, which, as we
have seen, are nothing but parameters $\mvec\theta$
of suitable models.  Using the alternative ways described in
Sects.~\ref{sect:reweighting} and \ref{sect:glob_inf}, we have
\begin{eqnarray}
 p(\mvec\theta \given \mvec d,\mvec h,I_0) &\propto&
\   p(\mvec d \given \mvec\theta,\mvec h,I_0) \,  p_0(\mvec\theta \given I_0) \\
 p(\mvec\theta \given \mvec d,I_0) &=& \int\! p(\mvec\theta \given \mvec d,\mvec h,I_0)
                     \,p(\mvec h \given I_0)\intd \mvec h
\label{eq:syst_gen_many}
\end{eqnarray}
and
\begin{eqnarray}
p(\mvec\theta,\mvec h \given \mvec d,I_0) &\propto& \  p(\mvec d \given  
\mvec\theta,\mvec h,I_0) \,
          p_0(\mvec\theta,\mvec h \given I_0)
\label{eq:globinf_many} \\
p(\mvec\theta|\,\mvec d,I_0) &=& \int\!p(\mvec\theta,\mvec h 
\given \mvec d,I_0)\intd \mvec h\,,
\label{eq:globinf_marg_many}
\end{eqnarray}
respectively. The two ways lead to an identical result, as it can be seen
comparing Eqs.~(\ref{eq:syst_gen_many}) and (\ref{eq:globinf_marg_many}).

Take a simple case of a common offset error
of an instrument used to measure various quantities $\mu_i$, 
resulting in the measurements $d_i$. We model each measurement as $\mu_i$
plus an error that is Gaussian distributed with a mean of zero and
a standard deviation $\sigma_i$. The
calculation of the posterior distribution can be performed analytically,
with the following results (see D'Agostini 1999c for details):
\begin{itemize}
\item
The uncertainty in each $\mu_i$ is described by a Gaussian centered
at $d_i$, with standard deviation $\sigma(\mu_i)=\sqrt{\sigma_i^2+\sigma_z^2}$,
consistent with Eq.~(\ref{eq:syst_z}).
\item
The joint posterior distribution $p(\mu_1,\mu_2,\ldots)$ does not factorize
into the product of $p(\mu_1)$, $p(\mu_2)$, etc., because correlations are
automatically introduced by the formalism, consistent with the
intuitive thinking of what a common systematic should do.
Therefore, the joint distribution will be a multi-variate Gaussian that
takes into account correlation terms.
\item
The {\it correlation coefficient} between any pair $\{\mu_i,\mu_j\}$
is given by
\begin{eqnarray}
\rho(\mu_i,\mu_j) &=& \frac{\sigma_z^2}{\sigma(\mu_i)\,\sigma(\mu_j)}
     = \frac{\sigma_z^2} 
{\sqrt{(\sigma_{i}^2+\sigma_z^2)\,(\sigma_j^2+\sigma_z^2)}}\,.
\end{eqnarray}
We see that $\rho(\mu_i,\mu_j)$ has the behavior expected from a
 common offset error; it is  non-negative; it varies from
 practically zero, indicating negligible correlation, when 
($\sigma_z\ll  \sigma_i$), to unity ($\sigma_z\gg  \sigma_i$), 
when the offset error dominates.
\end{itemize}

\subsection{Approximate methods and standard propagation applied
to systematic errors}

When we have many uncertain influence factors and/or the model of
uncertainty is non-Gaussian, the analytic solution of Eq.~(\ref{eq:syst_gen}),
or Eqs.~(\ref{eq:globinf})--(\ref{eq:globinf_marg})
can be complicated, or not existing at all.
Then numeric or approximate methods are needed. The most powerful
numerical methods are based on {\it Monte Carlo} (MC) techniques
(see Sect.~\ref{sect:computational} for a short account).
This issue goes beyond the aim of this report. In a recent
comprehensive particle-physics paper by Ciuchini \etal (2001), these ideas have
been used to infer the fundamental parameters of the
Standard Model of particle physics, using all available
experimental information.

For routine use, a practical approximate method can be developed by
thinking of the value inferred for the expected value of
$\mvec h$ as a {\it raw} value, indicated with $\mu_R$, that is,
$\mu_R = \mu\left|_{ \mvec{\scriptstyle h} \, = \, 
\mbox{\footnotesize E}(\mvec{\scriptstyle h})}\right.$
(`raw' in the sense that it needs later to be `corrected'
for all possible value of $\mvec \theta$, as it will be clear in a while). 
The value of $\mu$, which depends on the possible values of 
$\mvec h$, can be seen as a function of  $\mu_R$ and $\mvec h$:
\begin{eqnarray}
\mu &=& f(\mu_R, \mvec h)\,. \label{eq:mufmur}
\end{eqnarray}
We have thus turned our inferential problem into a standard problem of
evaluation of the pdf of a function of variables, of
which are particularly known the formulae to obtain
approximate values for expectations and standard deviations
in the case of independent {\it input quantities}
(following the nomenclature of ISO 1993):
\begin{eqnarray}
\mbox{E}(\mu) &\approx &  f(\mbox{E}(\mu_R), \mbox{E}(\mvec h)) \\
\sigma^2(\mu) &\approx &
\left(\left.\frac{\partial f}{\partial \mu_R}\right|_{\mbox{\small E}(\mu_R), 
\mbox{\small E}(\mvec h)}
\right)^2\!\sigma^2(\mu_R) \nonumber \\
& & + \sum_i  \left(\left.\frac{\partial f}
 {\partial h_i}\right|_{\mbox{\small E}(\mu_R), 
\mbox{\small E}(\mvec h)}
          \right)^2\! \sigma^2(h_i) \, .
\label{eq:sigma_prop}
\end{eqnarray}
Extension to multi-dimensional problems and treatment of correlations
is straightforward (the well-known covariance matrix propagation)
and we refer to (D'Agostini and Raso 1999) for details. 
In particular, this reference contains approximate formulae
valid up to second order, which allow to take
into account relatively easily non linearities. 

\section{Comparison of models of different complexity}
\label{sect:modelcomp}

We have seen so far two typical inferential situations:
\begin{enumerate}
\item
Comparison of simple models (Sect.~\ref{sect:simplemodels}),
where by simple we mean that
the models do not depend on parameters to be tuned to
the experimental data.
\item
Parametric inference given a model, to which we have devoted
the last sections.
\end{enumerate}
A more complex situation arises when we have several models, each
of which might depend on several parameters.
For simplicity, let us consider
model $A$ with  $n_A$ parameters $\mvec \alpha$ and
model $B$  with  $n_B$ parameters $\mvec \beta$.
In principle, the same Bayesian reasoning seen previously holds:
\begin{eqnarray}
\frac{P(A \given \mbox{Data},I)}{P(B \given \mbox{Data},I)}
&=& \frac{P(\mbox{Data} \given A,I)}{P(\mbox{Data} \given B,I)}\,
    \frac{P(A \given I)}{P(B \given I)}\,, \label{eq:BayesAB}
\end{eqnarray}
but we have to remember that the probability of the data, given a
model, depends on the probability of the data, given a model and any
particular set
of parameters, weighted with the prior beliefs about parameters.
We can use the same decomposition formula
 (see Tab.~\ref{tab:fx}), already applied
in treating systematic errors
(Sect.~\ref{sect:systematic}):
\begin{eqnarray}
P(\mbox{Data} \given M,I)& =& \int\!P(\mbox{Data} \given M,\mvec \theta, I)
\,p(\mvec \theta \given I)\intd \mvec\theta \,,\label{eq:evidence}
\end{eqnarray}
with $M=A,B$ and $\mvec \theta = \mvec \alpha, \mvec\beta$.
In particular, the Bayes factor appearing in Eq.~(\ref{eq:BayesAB}) becomes
\begin{eqnarray}
 \frac{P(\mbox{Data} \given A,I)}{P(\mbox{Data} \given B,I)}
 &=&
\frac
{\int\!P(\mbox{Data} \given A,\mvec \alpha, I)
\,p(\mvec \alpha \given I)\intd \mvec\alpha}
{\int\!P(\mbox{Data} \given B,\mvec \beta, I)
\,p(\mvec \beta \given I)\intd \mvec\beta}
 \label{eq:BayesfAB} \\
 &=& \frac
{\int\!{\cal L}_A(\mvec \alpha;\, \mbox{Data}) \,p_0(\mvec \alpha) \intd \mvec\alpha}
{\int\!{\cal L}_B(\mvec \beta;\, \mbox{Data}) \,p_0(\mvec \beta) \intd \mvec\beta}
\label{eq:IntLikeAB}\,.
\end{eqnarray}
The inference depends on the {\it marginalized likelihood}
(\ref{eq:evidence}), also known as the {\it evidence}.
Note that ${\cal L}_M(\mvec \theta;\, \mbox{Data})$
has its largest value around the maximum likelihood point
$\mvec \theta_{ML}$, but the evidence takes into account all
prior possibilities
of the parameters. Thus, it is not enough that the best fit of
one model is superior to its alternative, in the sense that, for instance,
\begin{eqnarray}
{\cal L}_A(\mvec \alpha_{ML};\, \mbox{Data})  &> &
{\cal L}_B(\mvec \beta_{ML};\, \mbox{Data})\,,
\end{eqnarray}
and hence, assuming Gaussian models,
\begin{eqnarray}
\chi^2_A(\mvec \alpha_{min\,\chi^2};\, \mbox{Data})  &< &
\chi^2_B(\mvec \beta_{min\,\chi^2};\, \mbox{Data})\,,
\end{eqnarray}
to prefer model $A$. We have already seen that we need to take
into account the prior beliefs in $A$ and $B$. But even this is not enough:
we also need to consider the space of possibilities and then
the adaptation capability of each model. It is well understood that
we do not choose an $(n-1)$ order polynomial as the best description
-- `best' in inferential terms --
of  $n$ experimental points,
 though such a model
always offers an exact  pointwise  fit. 
Similarly, we are much more impressed by,
and we tend {\it a posteriori} to believe more in,
a theory that absolutely predicts
an experimental observation, within a reasonable error,
than another theory that performs similarly or even better
after having adjusted many  parameters.

This intuitive reasoning
is expressed formally in Eqs.~(\ref{eq:BayesfAB}) and (\ref{eq:IntLikeAB}).
The evidence is given integrating the product
${\cal L}(\mvec \theta)$ and $p_0(\mvec \theta)$ over
the parameter space. So, the more
$p_0(\mvec \theta)$ is concentrated around $\mvec\theta_{ML}$,
the greater is the evidence in favor of that model. Instead,
a model with a volume of the parameter space much larger
than the one selected by ${\cal L}(\mvec \theta)$ gets disfavored.
The extreme limit is that of a  hypothetical model with so many
parameters to describe whatever we shall observe.
This effect is very welcome, and follows the {\it Ockham's Razor}
scientific rule of discarding unnecessarily complicated models
({\it ``entities should not be multiplied unnecessarily''}).
This rule comes out of the Bayesian approach automatically
and it is discussed, with examples of applications
in many papers.
Berger and Jefferys (1992)
introduce the connection between Ockham's Razor and Bayesian
reasoning, and discuss the evidence provided by the motion
of Mercury's perihelion in favor of Einstein's general relativity theory,
compared to alternatives at that time. Examples of recent applications are
Loredo and Lamb 2002 (analysis of neutrinos observed from 
supernova SN\ 1987A),   
John and Narlikar 2002 (comparisons of cosmological models),
Hobson \etal 2002 
(combination of cosmological datasets)
and Astone  \etal 2003 (analysis of 
coincidence data from gravitational wave detectors).
 These papers  also give
a concise account of underlying Bayesian ideas.

After having emphasized the merits of model comparison
formalized in Eqs.~(\ref{eq:BayesfAB}) and (\ref{eq:IntLikeAB}),
it is important to mention a related problem.
In parametric inference we have seen that we can make an
easy use of improper priors 
(see Tab.~\ref{tab:fx}), seen as limits of proper priors, essentially
because they simplify  in the Bayes formula. For example,
we considered $p_0(\mu \given I)$ of  Eq.~(\ref{eq:infGaussian})
to be a constant, but this constant goes to zero as the
range of $\mu$ diverges. Therefore, it does simplify in
 Eq.~(\ref{eq:infGaussian}), but not, in general, in
Eqs.~(\ref{eq:BayesfAB}) and (\ref{eq:IntLikeAB}), unless
models $A$ and $B$ depend on the same number of parameters
defined in the same ranges. Therefore, the general case
of model comparison is limited to proper priors, and needs
to be thought through better than when making
parametric inference.

\section{Choice of priors -- a closer look}
\label{sect:priors}

So far, we have considered mainly likelihood-dominated situations,
in which the prior pdf can be included in the normalization
constant.
But one should be careful about the possibility of
uncritically use uniform priors, as a `prescription,' 
or as a rule, though the rule might be associated with
the name of famous persons.
For instance, having made $N$ interviews
 to infer the proportion $\theta$ of a population
that supports a party, it is not reasonable to assume
a uniform prior of $\theta$ between 0 and 1. 
Similarly, having to infer the rate $r$
of a Poisson process (such that $\lambda = r \,T$, where $T$ is the
measuring time) related, for example, to proton decay, 
cosmic ray events or gravitational wave signals, 
we do not believe, strictly,
that $p(r)$ is uniform between zero and infinity. Besides
natural physical cut-off's (for example, very large
proton decay $r$ would prevent Life, or even stars,  to exist),
 $p(r) = \mbox{constant}$ implies to believe more high orders of magnitudes
of $r$ (see Astone and D'Agostini 1999 for details). In many cases 
(for example the mentioned searches for rare phenomena) our 
uncertainty could mean indifference over several orders of magnitude in
the rate $r$. This indifference can be parametrized roughly with a
 prior uniform $\ln r$ yielding $p(r) \propto 1/r$ 
(the same prior is obtainable using invariant arguments, as 
it will be shown in a while). 

As the reader might imagine, the choice of priors
is a highly debated issue, also among Bayesians.
We do not pretend to give definitive statements,
but would just like to touch on some important issues
concerning priors.

\subsection{Logical and practical role of priors}

Priors are pointed to by those critical of the Bayesian approach
as {\it the} major weakness of the theory.
Instead, Bayesians consider them 
a crucial and powerful key point of the method.
Priors are logically crucial because they are necessary to make
probability inversions via Bayes' theorem. This point remains valid
even in the case in which
they are vague and apparently disappear in the Bayes' formula. 
Priors are powerful because they allow to deal
with realistic situations in which informative prior knowledge
can be taken into account 
and properly balanced with the experimental information.

Indeed, we think that one of the advantages of Bayesian analysis 
is that it explicitly admits the existence of prior information, 
which naturally leads to the expectation that the prior will 
be specified  in any honest account of a Bayesian analysis. 
This crucial point is often obscured in other types of analyses,
in large part because the analysts maintain their method is `objective.' 
Therefore, it is not easy, in those analyses, to recognize
what are the specific assumptions made by the analyst 
--- in practice the analyst's priors --- and the assumptions included
in the method
(the latter assumptions are often unknown to the average practitioner).

\subsection{Purely subjective assessment of prior probabilities}
\label{sec:purely_subjective}

In principle, the point is simple, at least in 
one-dimensional problem in which there is good perception
of the possible range in which the uncertain variable of interest could
lie:\ try your best to model your prior beliefs. 
In practice, this advice seems difficult to follow because,
even if we have a rough idea of what the value of a quantity
should be, the representation of the prior in mathematical terms
seems very committal, 
because a pdf implicitly contains an infinite number
of precise probabilistic statements. (Even the uniform distribution says that we
believe {\it exactly} in the same way to all values. Who believes exactly that?)
It is then important to understand that, when expressing priors,
what matters is not the precise mathematical formula, 
but the gross value of the probability mass indicated by the formula, 
how probabilities are 
intuitively perceived and how priors influence posteriors.
When we say, intuitively, we believe something with a 95\% confidence, 
it means ``we are almost sure,'' but the precise value (95\%, instead of 92\%
or 98\%) is not very relevant. Similarly, when we
say that the prior knowledge is modeled by a Gaussian distribution
centered around $\mu_0$ with standard deviation $\sigma_0$
[Eq.~(\ref{eq:Gaussian_prior})], it means 
means that we are quite confident that $\mu$ is within
$\pm \sigma_0$, very sure that it is within
$\pm 2\sigma_0$ and {\it almost certain} that it is
within  $\pm $3\,$\sigma_0$. Values even farther from
$\mu_0$ are possible, though we do not consider them very likely.
But all models should be taken with a grain of salt, 
remembering that they
are often just mathematical conveniences. For example, 
a textbook-Gaussian prior includes infinite deviations from the expected value
and even negative values for physical quantities positively defined, like
a temperature or a length. All absurdities, if taken literally.
On the other hand, we think that all experienced physicists 
have in mind priors with low probability long tails in order to accommodate 
strong deviation from what is expected with highest probability.
(Remember that where the prior is zero, the posterior must also be zero.)

Summing up this point, it is important
to understand that
a prior should tell where the
{\it probability mass} is concentrated, without taking too seriously
the details, especially the tails of the distribution
(which should be, however, enough extended to accommodate 'surprises'). 
The nice feature of Bayes' theorem is the ability of trasform
such vague, fuzzy priors into solid estimates, if a sufficient amount
of good quality data are at hand. 
For this reason, the use of {\it improper priors} is not considered to be
problematic. Indeed, improper priors 
can just be considered a convenient way of modelling relative
beliefs. 
 
In the case we have doubts about the choice of the prior, we can 
consider a family of functions with some hyperparameters.
If we worry about the effect of the chosen prior
on the posterior, we can perform a
{\it sensitivity analysis}, i.e. to repeat the 
analysis for different, {\it reasonable} choices of the 
prior and check the variation of the result. 
The final uncertainty could, then, take into account
also the uncertainty on the prior. Finally, in extreme cases in which
priors play a crucial role and could dramatically change the conclusions,
one should refrain to give probabilistic result, providing, instead, only 
Bayes factors, or even just likelihoods. An example of a recent
result about  gravitational wave searches presented in this way, see
Astone \etal (2002). 

Having clarified meaning and role of priors, it is rather evident
that the practical choice of a prior depends on what is appropriate 
for the application.
For example, in the area of imaging, 
smoothness of a reconstructed image might be appropriate in 
some situations.  Smoothness may be imposed by a variety 
of means, for example, by simply setting the logarithm of 
the prior equal to an integral of the square of the second derivative 
of the image (von der Linden \etal 1996b). 
A more sophisticated approach goes under the name of Markov random 
fields (MRF),  which can even preserve sharp edges in the estimated images 
(Bouman and Sauer 1993, Saquib \etal 1997). A similar kind of prior 
is often appropriate for defomable geometric models, which can be used 
to represent the boundaries between various regions, for example, 
organs in medical images (Cunningham \etal 1998).

A procedure that helps in choosing the prior, expecially important
in the cases in which the
 parameters do not have a straightforwardly perceptible
influence on data, is to build a {\it prior predictive} pdf and
check if this pdf would produce data conform with our prior beliefs. 
The prior predictive distribution is the analogue of the 
({\it posterior}) predictive distribution we met in 
Sect.~\ref{sec:predictive}, with $p(\mvec \theta\given \mvec d, I)$ 
replaced by $p(\mvec \theta\given I)$ (note that the example of 
Sect.~\ref{sec:predictive} was one-dimensional, with $d_1=d_f$ and $\theta_1=\mu$), 
i.e. $p(\mvec d\given I) = \int p(\mvec d\given \mvec\theta, I)
\, p(\mvec\theta\given I)\intd\mvec\theta$.

Often, expecially in complicated data analyses, we are not sufficiently
knowledgable about the details of the problem. Thus, informative priors
have to be modelled that capture the judgement of experts. For
example, Meyer and Booker (2001) show a formal process of 
prior elicitation which has the aim at reducing, 
as much as possible, the bias in the experts' estimates 
of their confidence limits.  
This approach allows one to combine the results from several experts.
In short, we can suggest the use of the 
`coherent bet' (Sect.~\ref{sect:probability})
 to force experts to access
their values of probability, asking them to provide an interval in which
they feel `practically sure', 
intervals on which they could wager 1:1, and so on. 

\subsection{Conjugate priors}
\label{sect:conjugate}

Because of computational problems, 
modelling priors has been traditionally 
a compromise between a realistic assessment of beliefs
and choosing a mathematical function that simplifies 
the analytic calculations.
A well-known strategy is to choose a prior with a suitable form 
so the posterior belongs to the same functional family as the prior.
The choice of the family depends on the likelihood. A prior
and posterior chosen in this way are said to be {\it conjugate}.
For instance, given a Gaussian likelihood and choosing a Gaussian prior,
the posterior is still Gaussian, as we have seen in
Eqs.~(\ref{eq:Gaussian}), (\ref{eq:Gaussian_prior}) and
(\ref{eq:Gaussian_prior_inf}). This is because expressions of the form
$$
K\,\exp\left[ -\frac{(x_1-\mu)^2}{2\sigma_1^2}
           -\frac{(x_2-\mu)^2}{2\sigma_2^2}\right]
$$
can always be written in the form
$$
K'\,\exp\left[ -\frac{(x'-\mu)^2}{2\sigma'^2} \right] \, ,
$$
with suitable values for $x'$, $\sigma'$ and $K'$. The Gaussian distribution
is auto-conjugate. The mathematics is simplified but, unfortunately,
only one shape is possible.

An interesting case, both for flexibility and practical interest is
offered by the binomial likelihood (see Sect.~\ref{sect:binomial}).
Apart from the binomial coefficient, $p(n \given \theta, N)$ has the
shape $\theta^{n}(1-\theta)^{N-n}$, which has the same structure as the
{\it Beta distribution}, well known to statisticians: 
\begin{equation}
\mbox{Beta}(\theta \given r,s)=\frac{1}{\beta(r,s)}\theta^{r-1}(1-\theta)^{s-1}
\hspace{0.9cm}\left\{\!\begin{array}{l} 0\le \theta\le 1 \\
                               r,\,s > 0 \end{array}\right.\,,
\label{eq:distr_beta}
\end{equation}
where $\beta(r,s)$ stands for the Beta function, defined as
\begin{equation}
\beta(r,s)=\int_0^1 \theta^{r-1}(1-\theta)^{s-1}\intd \theta \,  
\end{equation}
which can be expressed in terms of Euler's Gamma function as
$\beta(r,s) =\Gamma(r)\,\Gamma(s)/\Gamma(r+s)$.
Expectation
and variance of the Beta distribution are:
\begin{eqnarray}
\mbox{E}(\theta)&=&\frac{r}{r+s} \label{eq:Ebeta}\\
\sigma^2(\theta)&=&\frac{r \, s}{(r+s+1)(r+s)^2}
= \mbox{E}^2(\theta)\, \frac{s}{r}\, \frac{1}{r+s+1} \,. \label{eq:Varbeta}
\end{eqnarray}
If $r>1$ and $s>1$, then the mode is unique, and it is at 
 $\theta_m = (r-1)/(r+s-2)$.
Depending on the value of the parameters the Beta pdf
can take a large variety of shapes.
For example, for large values of $r$ and $s$,
the function is very similar to a Gaussian distribution, 
while a constant function is obtained
for $r=s=1$.  Using the Beta pdf  as prior function in
inferential problems with a binomial likelihood, we have
\begin{eqnarray}
p(\theta \given n,N,r,s) &\propto &
    \left[\theta^n (1-\theta)^{N-n}\right] \left[\theta^{r-1}(1-\theta)^{s-1}\right] \\
    &\propto & \theta^{n+r-1} (1-\theta)^{N-n+s-1}\,.
\end{eqnarray}
The posterior distribution is still a Beta with
$r' = r+n$ and $s' =s+ N-n$, and expectation  and standard
deviation can be calculated easily from Eqs.~(\ref{eq:Ebeta})
and (\ref{eq:Varbeta}).
These formulae demonstrate how the posterior estimates become
progressively independent of the prior information in the limit of
large numbers;
this happens when both $m\gg r$ and $n-m\gg s$. In this limit, we get
the same result as for a uniform prior ($r=s=1$).
\begin{center}
\begin{table}
\caption{\sl Some useful conjugate priors.
$x$ and $n$ stand for the observed value (continuous or discrete, respectively) 
and $\theta$ is the generic
symbol for the
parameter to infer, corresponding to $\mu$ of a Gaussian, $\theta$ of a binomial
and $\lambda$ of a Poisson distribution.\vspace{2ex}}
\begin{tabular}{lll}\hline
likelihood & conjugate prior & posterior \\
$p(x \given \theta)$ &  $p_0(\theta)$ & $p(\theta \given x)$ \\
\hline
Normal$(\theta,\sigma)$ &  Normal$(\mu_0,\sigma_0)$ &
  Normal$(\mu_1,\sigma_1)$ \ \ [Eqs.~(\ref{eq:mu_average})--(\ref{eq:sigma_average})] \\
 Binomial$(N,\theta)$ & Beta$(r,s)$ & Beta$(r+n,s+N-n)$ \\
Poisson$(\theta)$ & Gamma$(r,s)$ & Gamma$(r+n,s+1)$ \\
Multinomial$(\theta_1,\ldots,\theta_k)$ &  Dirichlet$(\alpha_1,\ldots,\alpha_k)$ &
 Dirichlet$(\alpha_1+n_1,\ldots,\alpha_k+n_k)$ \\
\hline
\end{tabular}
\label{tab:conjugates}
\end{table}
\end{center}
Table~\ref{tab:conjugates} lists some of the more useful conjugate priors.
For a more complete collection of conjugate priors, 
see e.g. (Bernardo and Smith 1994, Gelman \etal 1995).

\subsection{General principle based priors}

Many who advocate using the Bayesian approach
still want to keep `subjective'
contributions to the inference to a minimum. 
Their aim is to derive prior functions based on
`objective' arguments or general principles.
As the reader might guess, this subject is rather
controversial, and the risk of trasforming 
arguments, which might well be reasonable and useful 
in many circumstances, into dogmatic rules is 
high (D'Agostini, 1999e).

\subsubsection{Transformation invariance}
\label{sect:invariance}
\label{sect:location} 
An important class of priors arises from the requirement 
of transformation invariance. 
We shall consider two specific cases, translation invariance 
and scale invariance. 
\begin{description}
\item[{\it Translation invariance}] 
\mbox{ }\\
Let us assume we are indifferent over a transformation
of the kind   $\theta' = \theta\,+\,b$, where $\theta$ is our 
variable of interest
and $b$ a constant. 
Then $p(\theta)\,\intd \theta$ is an infinitesimal mass element of probability 
for $\theta$ to be in the interval $\intd \theta$. Translation invariance
requires that this mass element remains unchanged when expressed 
in terms of $\theta'$, i.e. 
\begin{eqnarray}
p(\theta) \, \intd \theta &=& p(\theta') \, \intd \theta'  \label{eq:transl1} \\
                    &=& p(\theta + b) \, \intd \theta \,,  \label{eq:transl2}
\end{eqnarray}
since $\intd \theta = \intd \theta'$.
It is easy to see that in order for Eq.~(\ref{eq:transl2}) to hold for any $b$, 
$p(\theta)$ must be equal to a constant for all values of 
$\theta$ from $-\infty$ to $+\infty$. It is therefore an improper prior. 
 As discussed above, 
this is just a convenient modelling. For practical purposes this prior
should always be regarded as the limit for 
$\Delta\theta\rightarrow \infty$ of $p(\theta) = 1/\Delta \theta$,
where $\Delta \theta$ is a large finite range around the values of interest.

\item[{\it Scale invariance}]
\mbox{ }\\
In other cases, we could be indifferent about a scale transformation, 
that is  $\theta' = \beta \,\theta$, where $\beta$ is a constant. This invariance 
implies, since $\intd \theta' = \beta \intd \theta$ in this case,
\begin{eqnarray}
p(\theta) \, \intd \theta &=& p(\beta\, \theta) \, \beta \intd \theta \,,
\end{eqnarray}
i.e. 
\begin{eqnarray}
p(\beta\, \theta) &=& \frac{p(\theta)}{\beta}\,.
\label{eq:eqJeffreys}
\end{eqnarray}
The solution of this functional equation is
\begin{eqnarray}
p(\theta) &\propto&\frac{1}{\theta}\,, \label{eq:Jeffreys}
\end{eqnarray}
as can be easily proved using Eq.~(\ref{eq:Jeffreys}) as test solution in 
Eq.~(\ref{eq:eqJeffreys}). 
This is the famous {\it Jeffreys' prior}, since it was first proposed
by Jeffreys. Note that this prior also can be stated as 
$p(\log \theta) = \mbox{constant}$, as  can be easily verified. 
The requirement of scale invariance also 
produces an improper prior, in the range $0 < \theta <\infty$.
Again, the improper prior must be understood as the limit of a proper prior
extending several orders of magnitude around the values of interest.
[Note that we constrain $\theta$ to be positive because, traditionally,
variables which are believed to satisfy this invariance are associated with
positively defined quantities. Indeed, Eq.~(\ref{eq:Jeffreys}) 
has a symmetric solution for negative quantities.]
\end{description}
According to the supporters of these invariance motivated priors
(see e.g. Jaynes 1968, 1973, 1998, Sivia 1997, and  Fr\"ohner 2000, 
Dose 2002) variables associated to translation invariance are 
{\it location parameters}, as the parameter $\mu$ in a Gaussian model;
 variables associated to scale invariance are {\it scale parameters}, 
like $\sigma$ in a Gaussian model or $\lambda$ in a Poisson model.
For criticism about the (mis-)use of this kind 
of prior see (D'Agostini 1999d).

\subsubsection{Maximum-entropy priors}
\label{sect:maxent} 
Another principle-based approach to assigning priors 
is based on in the {\it Maximum Entropy principle} 
(Jaynes 1957a, also 1983, 1998, Tribus 1969, von der Linden 1995, Sivia
1997, and  Fr\"ohner 2000). The basic idea is to choose the prior 
function that maximizes the
Shannon-Jaynes {\it information entropy},
\begin{eqnarray}
S &=& - \sum_i^n p_i \, \ln p_i \, , \label{eq:entropy}
\end{eqnarray}
subject to whatever is assumed to be known about the distribution. 
The larger $S$ is, the greater
is our ignorance about the uncertain value of interest. 
The value $S=0$ is obtained
for a distribution that concentrates all the probability into a single value. 
In the case of no
constraint other than normalization, ($\sum_i^n p_i = 1$), 
$S$ is maximized by the uniform
distribution, $p_i = 1/n$, which is easily proved using Lagrange multipliers. 
For example, if the
variable is an integer between 0 and 10, a uniform distribution 
$p(x_i)=1/11$ 
gives $S=2.40$. 
Any binomial
distribution with $n=10$ gives a smaller value, with a maximum 
of $S=1.88$ for $\theta =1/2$ 
and a
limit of $S=0$ for $\theta \rightarrow 0$ or or $\theta \rightarrow 1$,
where $\theta$ is now the parameter of the binomial that gives
the probability of success at each trial.

Two famous cases of maximum-entropy priors for continuous variables
are when the only information about
the distribution is either the expected value or the expected 
value and the variance.
Indeed, these are special cases of general constraints
on the moments of the distribution $M_r$ (see Tab.~\ref{tab:fx}).
%
For $r=0$ and 1, $M_r$ is equal to 
unity and to the expected value, respectively. First and second moment 
together provide the variance (see Tab.~\ref{tab:fx} and 
Sect.~\ref{sec:MultidimensionalCase}). Let us sum up
what the assumed knowledge on the various moments provides 
[see e.g. (Sivia 1997, Dose 2002)]. 
\begin{description}
\item[{\it Knowledge about $M_0$}]
\mbox{ }\\
Normalization alone provides a uniform distribution over the interval 
in which the variable is defined: 
\begin{eqnarray}
p(\theta\given M_0=1) &=& \frac{1}{b-a} \hspace{0.6cm} a\le \theta \le b\,.
\end{eqnarray}
This is the extension to continuous variables
of the discrete case we saw above. 
\item[{\it Knowledge about $M_0$ and $M_1$  [i.e. about $\mbox{E}(\theta)$]}]
\mbox{ }\\
Adding to the constraint $M_0=1$ the knowledge about the 
expectation of the variable, plus
the requirement that all non-negative values are allowed, 
an exponential distribution is obtained: 
 \begin{eqnarray}
p(\theta\given M_0=1,M_1\equiv \mbox{E}(\theta)) &=& 
\frac{1}{\mbox{E}(\theta)}e^{-\theta/
 \mbox{\footnotesize E(}\theta\mbox{\footnotesize )}} 
\hspace{0.6cm} 0\le  \theta < \infty\,.
\end{eqnarray}
\item[{\it Knowledge about  $M_0$, $M_1$ and  $M_2$ 
[i.e. about $\mbox{E}(\theta)$ and $\sigma(\theta)$]}]
\mbox{ }\\
Finally, the further constraint provided by the standard deviation (related to 
first and second moment by the equation $\sigma^2 = M_2-M_1^2$) yields a prior
with a Gaussian shape independently of the range of $\theta$, i.e.
 \begin{eqnarray}
p(\theta\given M_0=1,\mbox{E}(\theta), \sigma(\theta)) &=& 
\frac{\exp\left[-\frac{(\theta-\mbox{\footnotesize E}(\theta))^2}
                      {2\, \sigma^2(\theta)} \right]}
{\int_a^b\exp\left[-\frac{(\theta-\mbox{\footnotesize E}(\theta))^2}
                         {2\, \sigma^2(\theta)} \right]\intd\theta }
\hspace{0.6cm}  a\le \theta \le b\,. \nonumber \\
\end{eqnarray}
The standard Gaussian is recovered when $\theta$ is 
allowed to be any real number.
\end{description}
Note, however, that
the counterpart of
Eq.~(\ref{eq:entropy}) for continuous variables is not trivial, since
all $p_i$ of Eq.~(\ref{eq:entropy}) tend to zero. 
Hence the analogous functional form 
$\int p(\theta) \ln p(\theta)\intd \theta$
no longer has a sensible
interpretation in terms of uncertainty, as remarked 
by Bernardo and Smith (1994). 
The Jaynes' solution is to introduce a `reference' 
density $m(\theta)$  to make entropy 
invariant under coordinate transformation via  
$\!\int p(\theta) \ln [p(\theta)/m(\theta)]\intd \theta$. 
(It is important to remark that the first and the 
third case discussed above are valid under the assumption 
of a unity reference density.) 
This solution is not universally
accepted (see Bernardo and Smith 1994), even though it conforms to the 
requirements of dimensional analysis.
Anyhow, besides formal aspects and the undeniable aid of 
Maximum Entropy methods in complicate problems such as
image reconstruction (Buck and Macauly 1991), we find it very
difficult, if not impossible at all, that a practitioner holds that
status of knowledge which give rise to the two celebrated
cases discussed above. We find more reasonable 
the approach described in Sect.~\ref{sec:purely_subjective},
that goes the other way around:
we have a rough idea of where the quantity of interest
could be, then we try to model it and to summarize it
in terms of expected value and standard deviation. In particular, we find 
untenable the position of those who state that 
Gaussian distribution can only be justified by Maximum Entropy
principle.

\subsubsection{Reference priors}
We conclude this discussion on priors by mentioning
 `reference analysis,'
which is an area of active research among statisticians.
The intention is, similarly to that for other priors motivated by basic principles,
that of ``characterizing a `{\it non-informative}' or
`{\it objective}' prior distribution, representing `{\it prior ignorance},'
`{\it vague prior knowledge},' and `{\it letting the data speak for themselves}' "
(Bernardo and Smith 1994). However, ``the problem is more complex
than the apparent intuitive immediacy of these words and phrases would suggest''
 (Bernardo and Smith 1994, p.~298):
\begin{quote}
{\sl \small
``Put bluntly:\ data cannot ever speak entirely for themselves: every 
prior specification
has {\it some} informative posterior or predictive implications; 
and `vague' is itself
much too vague an idea to be useful. There is no `objective' prior 
that represents
ignorance.

\hspace{1em} On the other hand, we recognize that there {\it is} 
often a pragmatically
important need for a form of prior to posterior analysis capturing,
{\it in some well-defined sense}, the notion of the prior having a minimal effect,
relative  to the data, on the final inference. Such a {\it reference analysis}
might be required as an approximation to actual beliefs; more typically, 
it might be
required as a limiting `what if?' baseline in considering a range of prior to
posterior analyses, or as a {\it default} option when there are insufficient
resources for detailed elicitation of actual prior knowledge.

\hspace{1em} \ldots From the approach we adopt, it will be clear 
that the {\it reference prior}
component of the analysis is simply a mathematical tool. It has considerable
pragmatic importance  in implementing a {\it reference analysis},
whose role and character will be precisely defined, but it is not a privileged,
`unique non-informative' or `objective' prior."
}
\end{quote}
The curious reader may take a look at the (Bernardo and Smith 1994) and 
references cited therein, as well as at Bernardo (1997).

\section{Computational issues}
\label{sect:computational}

The application of Bayesian ideas leads to computational problems, 
mostly related to the calculation of integrals for
normalizing the posterior pdf and for obtaining 
credibility regions, or simply the 
moments  of the distribution 
(and, hence, expectations, variances and covariances). 
The difficulties become challenging for problems involving many
parameters. This is one of the reasons why Bayesian inference 
was abandoned at the beginning of
the 20$^{\rm th}$ century in favor of simplified -- and simplistic -- methods. 
Indeed, the Bayesian renaissance over the past few decades is largely
due to the emergence of new numerical methods and the dramatic increases in
computational power, along  with clarifying work on the
foundations of the theory. 

\subsection{Overview of approximate computational strategies}
In previous sections we have already seen some `tricks' for simplifying
the calculations. The main topic of this section will be 
an introduction to Monte Carlo (MC). But, before doing that, we think
it is important to summarize the various `tricks' here. 
Much specialized literature is available on several aspects of 
computation in statistics.   
For an excellent review paper on the subject see (Smith 1991).
\begin{description}
\item[{\it Conjugate priors}] \mbox{ }\\
We discussed this topic in Sect.~\ref{sect:conjugate}, giving a couple 
of typical simple examples and references for a more detailed list of
famous conjugate distributions. We want to remark here that a conjugate prior
is a special case of the class of priors that simplify the calculation
of the posterior (the uniform prior is the simplest of this kind of prior). 
\item[{\it Gaussian approximation}] \mbox{ }\\
For reasons that are connected with the central limit theorem, 
when there is a large amount of consistent data the posterior 
tends to be Gaussian, practically independently of the exact shape 
of the prior. The (multi-variate) 
Gaussian approximation, which we encountered in 
Sect.~\ref{GaussianPosterior}, has an important role for applications, 
either as a reasonable approximation of the `true' posterior, or as 
a starting point for searching for a more accurate description of 
it. We also saw that in the case of practically flat priors
this method recovers the well-known minimum chi-square or maximum 
likelihood methods. 
\item[{\it Numerical integration}] \mbox{ }\\
In the case of low dimensional problems, standard numerical integration 
using either scientific library functions or the interactive tools of modern 
computer packages provide an easy solution to many problems (thanks
also to the graphical capabilities of modern programs which allow
the shape of the posterior to be inspected and 
the best calculation strategy decided upon). 
This is a vast and growing subject, into 
we cannot enter in any depth here, but we assume the 
reader is familiar with some of these programs or packages.  
\item[{\it Monte Carlo methods}] \mbox{ }\\
Monte Carlo methodology is a science in itself and it is 
way beyond our remit 
to provide an exhaustive introduction to it here. 
Nevertheless, we would like to introduce briefly some 'modern'
 (though the seminal work is already half a century old) methods
which are becoming extremely popular and are often associated 
with Bayesian analysis, the so called 
{\it Markov Chain Monte Carlo} ({\it MCMC}) methods. 
\end{description}

\subsection{Monte Carlo methods}
\subsubsection{Estimating expectations by sampling} 
\mbox{ } \\
The easy part of the Bayesian approach is to
write down the 
un-normalized distribution of the parameters
(Sect.~\ref{GaussianPosterior}), given the prior and the likelihood. 
This is simply $\tilde p(\mvec\theta\given \mvec d,I) 
= p(\mvec d\given \mvec\theta,I)\,  p(\mvec\theta\given I) $. 
The difficult task is to normalize this function and to calculate all
expectations in which we are interested, such as
expected values, variances, covariances
and other moments. We might also want to get marginal distributions, 
credibility intervals (or hypervolumes) and so on. 
As is well-known, if we were able to {\it sample} the posterior 
(even the un-normalized one),
i.e. to generate points of the parameter space according to their probability, 
we would have solved the problem, at least 
approximately. 
For example, the one-dimensional histogram of parameter $\theta_i$ 
would represent its marginal and would allow the calculation of
$\mbox{E}(\theta_i) \approx \left<\theta_i\right>$, 
$\sigma(\theta_i) \approx \left<\theta^2_i\right>-\left<\theta_i\right>^2$ and of
probability intervals ($\left<\cdot\right>$ in the previous formulae stand for arithmetic
averages of the MC sample). 

Let us consider the probability function $p(\mvec x)$ of 
the discrete variables $\mvec x$ and a function $f(\mvec x)$ 
of which we want to evaluate the expectation over the distribution 
$p(\mvec x)$. Extending the one-dimensional formula 
of Tab.~\ref{tab:fx}
to $n$ dimension
we have
\begin{eqnarray}
\mbox{E}[f(\mvec x)] &=& \sum_{x_1}\cdots\sum_{x_n} 
f(x_1,\ldots,x_n)\, p(x_1,\ldots,x_n) 
\label{eq:MCexpA} \\
                     &=& \sum_i f(\mvec x_i) \, p(\mvec x_i)\,,
\label{eq:MCexpB}
\end{eqnarray}    
where  the summation in Eq.~(\ref{eq:MCexpA}) is over the components, 
while  the summation in Eq.~(\ref{eq:MCexpB}) is over possible points
in the $n$-dimensional space of the variables. 
The result is the same. 

If we are able to sample a large number of points $N$ according to the 
probability function  $p(\mvec x)$, we expect 
each point to be generated $m_i$ times.
The average $\left<f(\mvec x)\right>$, calculated from the sample as 
\begin{eqnarray}
\left<f(\mvec x)\right> &=& \frac{1}{N} \sum_t   f(\mvec x_t)\,,
\label{eq:avovert}
\end{eqnarray} 
(in which the index is named $t$ as a reminder that this is a sum over 
a `time' sequence) can also be rewritten as 
\begin{eqnarray}
\left<f(\mvec x)\right> &=& \sum_i  f(\mvec x_i) \, \frac{m_i}{N}\,,
\end{eqnarray} 
just grouping together the outcomes giving the same $\mvec x_i$.
For a  very large $N$, the ratios $m_i/N$ are expected to be 
`very close' to $p(\mvec x_i)$ (Bernoulli's theorem),
and thus $\left<f(\mvec x)\right>$ becomes a good approximation 
of $\mbox{E}[f(\mvec x)]$. In fact, this approximation can be 
good (within tolerable errors) even if not all $m_i$ are large and, indeed, 
even if many of them are null.
Moreover, the same procedure can be
extended to the continuum, in which case `all points' ($\infty^n$)
can never be sampled.  

For simple distributions there are well-known standard techniques
for generating pseudo-random numbers starting from  pseudo-random numbers
distributed 
uniformly between 0 and 1 
(computer libraries are available for sampling 
points according to the most common
distributions). 
We shall not enter into 
these basic techniques, but will concentrate instead on 
the calculation of expectations in more complicated cases.

\subsubsection{Rejection sampling}
\mbox{ } \\
Let us assume we are able to generate points according to 
some function $g(\mvec x)$, such that, given a constant $c$, 
$p(\mvec x) \le cg(\mvec x)$. We generate $\mvec x^*$ according to 
$g(\mvec x)$ and decide to accept it with probability $p(\mvec x^*)/c g(\mvec x^*)$
(i.e. we extract another random number between 0 and 1 and accept the point 
if this number is below that ratio). 
It is easy to show that this procedure reshapes  $g(\mvec x)$ to
$p(\mvec x)$ and that it does not depend on the absolute normalization 
of $p(\mvec x)$
(any normalization constant can be absorbed in the multiplicative constant $c$).  
A trivial choice of $g(\mvec x)$, especially for simple one-dimensional problems,
is a uniform distribution (this variation is known as the {\it hit or miss} method), 
though clearly it can be  very inefficient. 

\subsubsection{Importance sampling}
\mbox{ } \\
In this method, too,  we start from an `easy' function $g(\mvec x)$, 
which `we hope' will approximate $p(\mvec x)$ of interest, of which
in fact we know only its un-normalized expression 
$\tilde p(\mvec x)$. However, there is no requirement about how $g(\mvec x)$
approximates $p(\mvec x)$ (but the goodness of approximation
will have an impact on the efficacy of the method), 
apart from the condition that $g(\mvec x_i)$ must be positive
wherever  $p(\mvec x_i)$ is positive . 

The function $g(\mvec x)$ can be used
in the calculation of $\mbox{E}[f(\mvec x)]$, if we 
notice that $\mbox{E}[f(\mvec x)]$ can be rewritten as follows:
\begin{eqnarray}
\mbox{E}[f(\mvec x)] &=& \frac{\int\! f(\mvec x) \, \tilde p(\mvec x)\intd \mvec x}
                             {\int\! \tilde p(\mvec x)\intd \mvec x} 
\label{eq:impsampl1} \\
&=& \frac{\int\! f(\mvec x) \, 
\left[\tilde p(\mvec x)/g(\mvec x)\right]\,g(\mvec x)\intd \mvec x}
{\int\! \left[\tilde p(\mvec x)/g(\mvec x)\right]\,g(\mvec x) \intd \mvec x} \\ 
&=&    \frac{\mbox{E}_g\left[ f(\mvec x) \, 
       \tilde p(\mvec x)/g(\mvec x) \right]}
         {\mbox{E}_g\left[\tilde p(\mvec x)/g(\mvec x)\right]}\,,
\end{eqnarray}
where the the symbol $\mbox{E}_g$ is a reminder that the expectation is
calculated  over
the distribution $g(\mvec x)$. Finally, the strategy can be implemented 
in the Monte Carlo using Eq.~(\ref{eq:avovert}) for the two expectations:
\begin{eqnarray}
\mbox{E}[f(\mvec x)] & \approx &
 \frac{ \sum_t f(\mvec x_t) \, \tilde p(\mvec x_t)/g(\mvec x_t)}
      { \sum_t \tilde p(\mvec x_t)/g(\mvec x_t) }\,.
\end{eqnarray}
From the same sample it is also possible to evaluate the normalization constant,
given by the denominator of Eq.~(\ref{eq:impsampl1}), i.e.
\begin{eqnarray}
Z &=&  \int\! \tilde p(\mvec x)\intd \mvec x \approx \frac{1}{N} 
\sum_t \frac{\tilde p(\mvec x_t)}{g(\mvec x_t)}\,.
\end{eqnarray}
The computation of this quantity is particularly important 
when we are dealing with model comparison and $Z$ has the meaning of `evidence'
(Sect.~\ref{sect:modelcomp}).  

It easily to see that the method works well if $g(\mvec x)$ overlaps well
with $p(\mvec x)$. Thus, a proper choice of $g(\mvec x)$ can be made 
by studying where the probability mass of $p(\mvec x)$ is concentrated
(for example finding the mode of the distribution in a numerical way). 
Often a Gaussian function is used for $g(\mvec x)$, with parameters
chosen to approximate  $p(\mvec x)$ in the proximity of the mode, 
as described in Sect.~\ref{GaussianPosterior}. In other cases, other functions
can be used which have more pronounced tails, like $t$-Student 
or Cauchy distributions. Special techniques, into which we cannot enter here, 
allow $n$ independent random numbers to be generated and, subsequently, 
by proper rotations, turned into other numbers which have a correlation matrix
equal to that of the multi-dimensional 
Gaussian which approximates $p(\mvec x)$.

Note, finally, that, contrary to the rejection sampling,
importance sampling is not suitable for generate samples of `unweighted events', 
such as those routinely used in the planning and the analysis
 of many kind experiments, 
especially particle physics experiments. 
 
\subsubsection{Metropolis algorithm}
A different class of Monte Carlo methods is based on Markov chains
and is known as Markov Chain Monte Carlo. 
The basic difference from the methods described above is that the 
sequence of generated points takes a kind of random walk in parameter space, 
instead of each point being generated, one independently from another. Moreover,
the probability of jumping from one point to an other depends only
on the last point and not on the entire previous history (this is the peculiar
property of a Markov chain). There are several MCMC algorithms.  
One of the  most popular and simple algorithms, 
applicable to a wide class of problems, is the
{\it Metropolis} algorithm (Metropolis \etal 1953).
One starts from an arbitrary point $\mvec x_0$ and generates the sequence
by repeating the following cycle, 
with $\mvec x_t$ being the previously selected point at each iteration:
\begin{enumerate}
\item
Select a new trial point $\mvec x^*$, chosen according to a 
symmetric {\it proposal} pdf
 $q(\mvec \theta^*\given \mvec \theta_t)$.
\item
Calculate the {\it acceptance probability} 
\begin{equation}
A(\mvec x^*\given \mvec x_t) = 
\mbox{min}\left[1, \,
\frac{\tilde p(\mvec \theta^*)}{\tilde p(\mvec \theta_t)} \right] \,.
\end{equation}
\item
Accept $\mvec x^*$ with probability $A(\mvec x_t,\mvec x^*)$, i.e.
\begin{itemize}
\item 
if  $\tilde p(\mvec \theta^*) \ge \tilde p(\mvec \theta_t)$, then accept $\mvec x^*$;
\item
if  $\tilde p(\mvec \theta^*) < \tilde p(\mvec \theta_t)$, extract a uniform random
number between 0 and 1 and accept  $\mvec x^*$ if 
the random number is less then $\tilde p(\mvec \theta^*) / \tilde p(\mvec \theta_t)$.
\end{itemize}
If the point is accepted, then $\mvec x_{t+1}= \mvec x^*$. Otherwise 
$\mvec x_{t+1}= \mvec x_t$
\end{enumerate}
This algorithm allows a jump $\mvec x_t$ to $\mvec x_{t+1}$ with probability
$T(\mvec x_{t+1}\given \mvec x_t)$ (the {\it transition kernel}) 
equal to $A(\mvec x^*\given \mvec x_t) \,q(\mvec \theta^*\given \mvec \theta_t)$. 
The algorithm has a stationary asymptotic distribution equal to 
$p(\mvec x)$ (i.e. the chain is {\it ergodic}) and ensures {\it detailed balance}:
\begin{equation}
p(\mvec  x_{t+1})\, T(\mvec x_{t}\given \mvec x_{t+1})
= p(\mvec  x_{t})\, T(\mvec x_{t+1}\given \mvec x_{t})\,.
\end{equation}
By construction, the algorithm does not depend on the 
normalization constant, since what matters is the ratio of the pdf's. 
The variation of the algorithm in which the proposal pdf $q()$ is not symmetric
is due to Hasting (1970) and for this reason the algorithm is 
often  also called
Metropolis-Hasting. Moreover, what has been described here is the 
{\it global} Metropolis algorithm, in contrast to the {\it local} one, 
in which a cycle affects only one component of $\mvec x$. 

The fact that this algorithm belongs to the class of MCMC 
gives rise to two problems. 
First, each point in the chain has some correlation with the points
which immediately preceded it,  
and usually the chain moves slowly (and irregularly) from one region 
in the  variable space to another
(note also that, if a proposed point is not accepted, 
the chain keep the same position in the next step, and this
circumstance can happen several times consecutively). 
Second, the initial part of the sequence is strongly 
influenced by the arbitrary starting point. 
Therefore, it is necessary to 
remove the initial part of the chain. 

Using an MCMC for a complex problem is not an automatic procedure
and some tuning is needed. One of the important things to choose with care is the 
proposal function. If too small jumps are proposed, the chain moves too slowly 
and, can even remain trapped in a subregion and never sample the 
rest of the parameter space, 
if the probability distribution is defined over disconnected regions. 
If too large steps are proposed, the proposed points could often fall in very
low probability regions and not be accepted, 
in which case the chain remains stuck
in a point for many cycles. 

For an interesting, insightful introduction to principles and applications 
of MCMC see (Kass \etal 1998). A nice tutorial is given 
by (Hanson 2000). A recent application of Bayesian methods in cosmology, which
uses MCMC and contains a pedagogical introduction too,
can be found in (Lewis and Bridle 2002). 
For a good treatise, freely available on the
web, (Neel 1993) is recommended. The reader will find that MCMC techniques
are used to solve complex problems graphically represented in
terms of {\it Bayesian networks} (also known as {\it belief networks}
or simply {\it probabilistic network}). This subject, which has 
revolutionized the way of thinking artificial intelligence and the
uncertainty issues related to it, does beyond the purpose of this article. 
The interested reader can find more information in (Pearl 1988, 
BUGS 1996, Cowell \etal 1999 and Cozman 2001) and references therein.

\section{Conclusions}
The gain in popularity Bayesian methods have enjoyed in recent years
is due to various conceptual and practical advantages
they have over other approaches, among which are: 
\begin{itemize}
\item
the recovery of the intuitive idea of probability as a valid 
concept for treating scientific problems;
\item
the simplicity and naturalness of the basic tool;
\item
the capability of combining prior knowledge and experimental
information;
\item
the property permitting automatic updating as soon as new 
information becomes available;
\item
the transparency of the methods, which allow the 
different assumptions upon which an inference may depend
to be checked and changed;
\item
the high degree of awareness the methods give to the user.
\end{itemize}
In this article we have seen how to build a theory 
of uncertainty in measurement as a straightforward application
of the basic Bayesian ideas, without unnecessary
principles or {\it ad hoc} prescriptions. In particular,
the uncertainty due to systematic errors can be treated
in a consistent and powerful way.

Providing an exact solution for inferential problems
can easily lead to computational difficulties. 
We have seen several ways to overcome such difficulties, 
either by using suitable approximations, or by using modern 
computational methods. In particular, it has been shown
that the approximate solution often coincides
with a `conventional' method, but only under well 
defined conditions. Thus, for example, 
minimum $\chi^2$ formulae can be used, with a Bayesian
spirit and with  a natural interpretation 
of the results,  in all those routine cases 
in which the analyst considers as reasonable the conditions 
of their validity.

A variety of examples of applications have been shown, or
mentioned, in this paper. Nevertheless, the aim of the author 
was not to provide a complete review of
Bayesian methods and applications, but rather to introduce
those Bayesian ideas that could be of help in understanding 
more specialized literature. 
Compendia of the Bayesian theory are
given in (Bernardo and Smith 1994, O'Hagan A 1994 and 
Robert 2001). Classic, influential books are
(Jeffreys 1961, de Finetti 1974, Jaynes 1998). 
Among the many books introducing Bayesian methods, 
(Sivia 1996) is particularly suitable for physicists. 
Other recommended texts which treat general aspects of data analysis are
(Box and Tiao 1973, Bretthorst 1988, 
Lee 1989, Gelman \etal 1995, 
Cowell \etal 1999, Denison \etal 2002, 
Press 2002). 
More specific applications can be found in the 
proceedings of the conference series and several web sites. 
Some useful starting points for web navigation are given: 
 \\ \\ 
\begin{tabular}{ll}
ISBA  book list      &  http://www.bayesian.org/books/books.html \\
UAI proceedings &  http://www2.sis.pitt.edu/~dsl/UAI/uai.html \\
BIPS            &  http://astrosun.tn.cornell.edu/staff/loredo/bayes/ \\
BLIP            &  http://www.ar-tiste.com/blip.html \\
IPP Bayesian analysis group   &  http://www.ipp.mpg.de/OP/Datenanalyse/ \\
Valencia meetings&  http://www.uv.es/$^{\footnotesize\sim}$bernardo/valenciam.html \\
Maximum Entropy resources&  http://omega.albany.edu:8008/maxent.html \\
MCMC preprint service & http://www.statslab.cam.ac.uk/$^{\footnotesize\sim}$mcmc/
\end{tabular}
 \\ \vspace{0.5cm} \\ 
I am indebted to Volker Dose and Ken Hanson for extensive 
discussions concerning the contents of this paper, as well as
for substantial editorial help. 
The manuscript has also benefited from comments
by Tom Loredo.

\newpage
\noindent {\bf References}\
\vspace{1ex}
\begin{itemize}
\normalsize

\item[] 
Astone P \etal 2002 Search for correlation 
between GRB's detected by BeppoSAX 
and gravitational wave detectors EXPLORER and NAUTILUS 
{\it Phys. Rev.} {\bf 66} 102002.
\item[] 
Astone P and D'Agostini G 1999 
Inferring the intensity of Poisson processes at limit 
of detector sensitivity (with a case study on gravitational wave burst search)
{\it CERN-EP/99-126}
\item[] 
Astone P, D'Agostini G and D'Antonio 2003
Bayesian model comparison applied to the 
Explorer-Nautilus 2001 coincidence data, 
arXiv:gr-qc/0304096
\item[] 
Babu G J and Feigelson E D  1992 eds 
{\it Statistical Challenges in Modern Astronomy I}  (New York:\ Springer)
\item[] 
Babu G J and Feigelson E D 1997 eds 
{\it Statistical Challenges in Modern Astronomy II}  (New York:\ Springer)
\item[] 
Berger J O and Jefferys W H 1992 
Sharpening Ockham's razor on a Bayesian strop
{\it Am. Scientist} {\bf 89} 64--72
  and {\it J. Ital. Stat. Soc.} {\bf 1} 17 
\item[] 
Bernardo J M 1999 ed {\it Bayesian Methods in the Sciences}, 
special issue of {\it Rev. Acad. Cien. Madrid}, {\bf 93}(3)
\item[] 
Bernardo J M, J O Berger, A P Dawid and A F M Smith 1999 eds
     {\it Bayesian Statistics~6} (Oxford: Oxford University)
\item[] 
Bernardo J M and Smith F M 1994 {\it Bayesian Theory} 
(Chichester:\ John Wiley \& Sons)
\item[] 
Bernardo J M 1997 Non-informative priors do not exist 
{\it J. Stat. Planning and Infer.} {\bf 65} 159 
\item[] 
Bontekoe T R, Koper E and Kester D J M 1994
 Pyramid maximum entropy images of IRAS survey data 
{\it Astron. Astrophys.} {\bf 284} 1037 
\item[] 
Bouman C A and Sauer K 1993 A generalized Gaussian image model 
for edge-preserving MAP estimation 
{\it IEEE Trans. on Image Processing} {\bf 2} 296 
\item[]
Box G E P and Tiao G C 1973
{\it Bayesian inference in statistical analysis}
(Chichester:\ J. Wiley \& Sons)
\item[]
Bretthorst G L 1988 {\it Bayesian spectrum analysis and parameter 
estimation} (Berlin: Springer)
\item[]
BUGS 1996 http://www.mrc-bsu.cam.ac.uk/bugs/welcome.shtml
\item[]
Buck B and Macauly V A eds 1991 {\it Maximum Entropy in action}, 
(Oxford: Oxford University Press)
\item[] 
Ciuchini M \etal 2001
2000 CKM-Triangle Analysis: A critical review 
with updated experimental inputs and theoretical parameters
 {\it J. High Energy Phys.} {\bf 0107} 013
\item[] 
Coletti G and Scozzafava R 2002
{\it Probabilistic logic in a coherent setting''}, 
 (Dordrecht: Kluwer Academic)
\item[] Cousins R D 1995 Why isn't every physicist a Bayesian? 
{\it Am. J. Phys.} {\bf 63} 398 
\item[]  
Cowell R G,  Dawid A P,  Lauritzen S L  and  Spiegelhalter D J 1999
{\it Probabilistic Networks and Expert Systems}, (New York:\ Springer)
\item[]
Cox R T 1946 Probability, Frequency and Reasonable Expectation
{\it Am. J. Phys.}  {\bf 14} 1 
\item[]
Cozman F B 2001 
JavaBayes version 0.346 -- Bayesian networks in Java 
http://www-2. cs.cmu.edu/$\sim$javabayes/Home/
\item[] 
Cunningham G\ S, Hanson K\ M and Battle X\ L 1998 
Three-dimensional reconstructions from low-count SPECT data 
using deformable models {\it Opt. Expr.} {\bf 2} 227 
\item[] 
D'Agostini\ G 1999a Bayesian Reasoning versus Conventional Statistics 
in High Energy Physics
{\it Maximum Entropy and Bayesian Methods}
        ed von der Linden W {\etal} (Dordrecht:\ Kluwer Academic)
\item[]
 D'Agostini\ G 1999b Sceptical combination of experimental results:
 General considerations and application to epsilon-prime/epsilon
{\it CERN-EP/99-139}
\item[] 
D'Agostini\ G 1999c {\it Bayesian reasoning in high-energy 
physics:\ principles and applications} {\it CERN Report 99-03}
(an extended version of this report is going to be published as 
{\it Bayesian reasoning in data analysis -- A critical introduction}
by World Scientific Publishing)
\item[] 
D'Agostini\ G 1999d Teaching statistics in the physics curriculum:\ 
Unifying and clarifying role of subjective probability  
{\it Am. J. Phys.}  {\bf 67}  1260  
\item[] 
D'Agostini\ G 1999e Overcoming prior Anxiety, 
{\it Bayesian Methods in the Sciences} ed 
J. M. Bernardo; 
special issue of {\it Rev. Acad. Cien. Madrid} {\bf 93}(3),
        311 
\item[] 
D'Agostini\ G 2000 Confidence limits: what is the problem? 
Is there {\it the} solution?
{\it CERN Report 2000-005}
 ed James F and Lyons L (Geneva:\ CERN) 3
\item[] 
D'Agostini\ G 2002 Minimum bias legacy of search results 2002
{\it Nucl Phys Proc Suppl} {\bf 109}  148  
\item[] 
D'Agostini G and Degrassi G 1999
Constraints on the Higgs Boson Mass from Direct 
Searches and Precision Measurements 
{\it Eur. Phys. J.} {\bf C10} 663 
\item[] 
D'Agostini G and Raso M 
Uncertainties due to imperfect knowledge of systematic effects: 
general considerations and approximate formulae
{\it CERN-EP/2000-026}
\item[] 
de Finetti B 1974 {\it Theory of Probability} 
(Chichester:\ J. Wiley \& Sons)
\item[]
Denison D G T, Holmes C C, Mallick B K and Smith A F M 2002
{\it Bayesian methods for nonlinear classification and regression}
(Cichester: J. Wiley \& Sons)
\item[] 
DIN (Deutsches Institut f\"ur Normung) 1996
{\it Grundlagen der Messtechnik - Teil 3: 
Auswertung von Messungen einer einzelnen Messgröße, Messunsicherheit} DIN 1319-3 
(Berlin: Beuth Verlag)
\item[] 
DIN (Deutsches Institut f\"ur Normung) 1999
{Grundlagen der Messtechnik - Teil 4: Auswertung von Messungen, Messunsicherheit}
DIN 1319-4 (Berlin: Beuth Verlag)
\item[] 
Dose V 2002 Bayes in five days, 
CIPS, MPI f\"ur Plasmaphysik, Garching, Germany, Reprint 83, May 2002
\item[] 
Dose V and von der Linden W 1999 Outlier tolerant parameter 
estimation {\it Maximum Entropy and
        Bayesian Methods} ed von der Linden W \etal (Dordrecht:\
        Kluwer Academic) 47 
\item[] 
Efron B  1986a Why isn't everyone a Bayesian?
{\it Am. Stat.} {\bf 40} 1
\item[] 
Efron B 1986b reply to Zellner 1986 {\it Am. Stat.} {\bf 40} 331
\item[] 
Fischer R, Mayer M, von der Linden W and Dose V 1997
 Enhancement of the energy resolution in ion-beam 
experiments with the maximum-entropy method 
{\it Phys.~Rev.~E} {\bf 55} 6667 
\item[] 
Fischer R, Mayer M, von der Linden W and Dose V 1998 
Energy resolution enhancement in ion beam experiments 
with Bayesian probability theory {\it Nucl.~Instr.~Meth.} {\bf 136-138}
1140 
\item[] 
Fischer R, Hanson K M, Dose V and von der Linden W 2000 
Background estimation in experimental spectra {\it Phys.~Rev.} {\bf E61} 1152 
\item[] 
Fr\"ohner F H 2000 {\it Evaluation and Analysis of Nuclear
Resonance Data} JEFF Report 18 (Paris: OECD Publications)
\item[] 
Gelman A, Carlin J B, Stern H S and Rubin D B 1995 {\it
Bayesian Data Analysis} (London:\ Chapman and Hall)
\item[] 
Glimm J and Sharp D H 1999 Prediction and the quantification of uncertainty 
{\it Physica D} {\bf 133} 152  
\item[] 
Gregory P C and Loredo T J 1992 
A new method for the detection of a periodic signal 
of unknown shape and period {\it Astr.~J.} {\bf 398} 146 
\item[] 
Gregory P C and Loredo T J 1996 Bayesian periodic signal detection II 
-- Bayesian periodic signal detection:\ analysis of ROSAT observations 
of PSR 0540-693 {\it Astr.~J.} {\bf 473} 1059  
\item[] 
Gregory P C 1999 
Bayesian periodic signal detection I --
 analysis of 20 years of radio flux measurements of 
the x-ray binary LS I +61$^\circ$303 
{\it Astr.~J.} {\bf 520} 361  
\item[] 
Gubernatis J E, Jarrell M, Silver R N and Sivia D S  1991 
Quantum Monte-Carlo simulations and maximum-entropy:\ dynamics 
from imaginary-time data {\it Phys. Rev. B} {\bf 44} 6011  
\item[] 
Hanson K M 1993 
Introduction to Bayesian image analysis
{\it Medical Imaging: Image Processing}  
Loew M H ed  {\it Proc. SPIE} {\bf 1898} 716 
\item[] 
Hanson K M 2000 Tutorial on Markov Chain Monte Carlo, 
{\it http://public.lanl.gov/ kmh/talks/maxent00b.pdf}
\item[]
Hasting W K 1970 Monte Carlo sampling methods using Markov chains and
their applications {\it Biometrica} {\bf 57} 97 
\item[]
Higdon D M and Yamamoto S Y 2001 
Estimation of the head sensitivity function 
in scanning magnetoresistance microscopy 
{\it J. Amer. Stat. Assoc.} {\bf 96} 785 
\item[]
 Hobson M P, Bridle S L and Lahav 2002
Combining cosmological datasets: 
hyperparameters and Bayesian evidence
 arXiv:astro-ph/0203259
\item[] 
Howson C and Urbach P 1993 
{\it Scientific reasoning --- the Bayesian approach} (Chicago and La Salle:  Open Court)
\item[] 
ISO (International Organization for Standardization) 1993
{\it Guide to the Expression of Uncertainty in Measurement} (Geneva:\ ISO)
\item[] 
Jaynes E T 1957a Information theory and statistical mechanics {\it Phys.~Rev.}
 {\bf 106} 620 
\item[] 
Jaynes E T 1957b Information theory and statistical mechanics II 
{\it Phys.~Rev.} {\bf 108} 171 
\item[] 
Jaynes E T 1968 Prior probabilities {\it IEEE Trans. Syst. Cybern.}
{\bf SSC-4} 227, reprinted in (Jaynes 1983)
\item[] 
Jaynes E T 1973 The well-posed problem
{\it Found. Phys.} {\bf 3} 477, reprinted in (Jaynes 1983)
\item[] 
Jaynes E T 1983 {\it Papers on Probability, Statistics
and Statistical Physics} ed Harper W L and Hooker C A (Dordrecht:\ Reidel)
\item[] 
Jaynes E T 1998 {\it http://omega.albany.edu:8008/JaynesBook.html}
\item[] 
Jeffreys H 1961  {\it Theory of Probability} (Oxford: Oxford University)
\item[] 
John M V and Narlikar J V 2002 
Comparison of cosmological models using Bayesian theory {\it Phys. Rev.} 
{\bf D65} 43506 
\item[] 
Kadane J B and Schum D A 1996 {\it A Probabilistic Analysis of the
Sacco and Vanzetti Evidence} (Chichester:\ Wiley and Sons)
\item[]
Kalman R E 1960 A new approach to linear filtering and prediction problems 
{\it Trans. ASME Journal of Casic Engineering} {\bf 82}  35 
\item[] 
Kass R E, Carlin B P, Gelman A and Neal R M
 1998 Markov Chain Monte Carlo in practice: A roundtable discussion 
{\it Am.\ Stat.} {\bf 52} 93  
\item[] 
Lad F 1996  {\it Operational Subjective Statistical Methods --
a Mathematical, Philosophical, and Historical Introduction} 
(Chichester:J. Wiley \& Sons)
\item[] 
Lee P M 1989 {\it Bayesian statistics - an introduction} 
(Chichester:J. Wiley \& Sons)
\item[] 
Lewis A and Bridle S 2002
Cosmological parameters from CMB and other data:
a Monte-Carlo approach {\it  Phys.Rev.} {\bf D66} 103511
\item[] 
von der Linden W 1995 Maximum-entropy data analysis 
{\it Appl.~Phys.} {\bf A60} 155 
\item[] 
von der Linden  W, Dose V and Fischer R 1996b 
Spline-based adaptive resolution image reconstruction 
{\it Proceedings of the 1996 Maximum Entropy Conference}
ed Sears M \etal 
(Port Elizabeth:\ N.M.B.~Printers) 154 
\item[] 
Lindley D V 1986 Discussion to Efron 1986a {\it Am. Stat.} {\bf 40} 6
\item[]
Loredo T J  1990  {\it Maximum Entropy and Bayesian Methods}
ed Foug\'{e}re P F (Dordrecht:\ Kluwer Academic) 81
\item[]
Loredo T J and Lamb D Q 2002 
Bayesian analysis of neutrinos observed from supernova SN 1987A
{\it Phys. Rev.} {\bf D65} 063002
\item[] 
Malakoff D 1999 Bayes Offers a 'New' Way to Make Sense of Numbers
{\it Science} {\bf 286} 1460
\item[] 
Maybaeck P S 1979 {\it Stochastic models, estimation and control}, 
Vol. 1 (New York:\ Academic Press).  
\item[]
Metropolis  H, Rosenbluth A W, Rosenbluth M N, Teller A H and 
Teller E 1953 Equations of state calculations by fast computing
machines {\it Journal of Chemical Physics} {\bf 21} 1087
\item[] 
von Mises R 1957 {\it Probability, Statistics, and Truth} 
(St Leonards:\ Allen and Unwin); reprinted in 1987 by Dover
\item[]
Neal R M 1993 {\it Probabilistic inference using 
Markov chain Monte Carlo methods} (Toronto: Technical Report CRG-TR-93-1)
\item[]
O'Hagan A 1994 {\it Kendall's Advanced Theory of 
Statistics:\ Vol.\ 2B - Bayesian Inference} (New York:\ Halsted)
\item[] 
Pearl J 1988  
{\it Probabilistic Reasoning in Intelligent Systems:\
Networks of Plausible Inference}
(San Mateo:\ Morgan Kaufmann)
\item[] 
Press W H 1997 Understanding data better with 
Bayesian and global statistical methods 
{\it Unsolved problems in astrophysics}49--60
        ed Bahcall J~N and Ostriker J~P  
(Princeton:\ Princeton University) 49 
\item[]
Press S J 2002 {\it Subjective and Objective Bayesian Statistics: 
Principles, Models, and Applications} 2nd Edition
(Chichester:\ John Wiley \& Sons)
\item[] 
Robert C P 2001 {\it The Bayesian Choice} (New York:\ Springer)
\item[] 
Saquib S S, Hanson K M, and Cunningham G S 1997 
Model-based image reconstruction from time-resolved diffusion data 
{\it Proc. SPIE} {\bf 3034} 369  
\item[] 
Schr\"odinger E 1947a The Foundation of the Theory of Probability -- I 
{\it Proc. R. Irish Acad.} 
{\bf 51A} 51; 
reprinted in {\it Collected papers} Vol. 1 
(Vienna 1984: Austrian Academy of Science) 463 
\item[] 
Schr\"odinger E 1947b The Foundation of the Theory of Probability -- II 
{\it Proc. R. Irish Acad.} 
{\bf 51A} 141; 
reprinted in {\it Collected papers} Vol. 1 
(Vienna 1984: Austrian Academy of Science) 479 
\item[] 
Sivia D S 1997  {\it Data Analysis -- a Bayesian Tutorial}
(Oxford:\ Clarendon)
\item[] 
Skilling J 1992 Quantified maximum entropy 
{\it Int.~Spectr.~Lab.} {\bf 2} 4 
\item[]
 Smith A F M 1991 Bayesian numerical analysis 
{\it Phyl. Trans. R. Soc. London} {\bf 337} 369 
\item[] Taylor B N and  Kuyatt C E 1994
{\it Guidelines for Evaluating and Expressing Uncertainty of 
NIST Measurement Results}
(Gaithersburg: NIST Technical Note 1297); 
available on line at {\it http://physics.nist.gov/}
\item[] 
Tribus M 1969 {\it Rational Descriptions, Decisions, and Designs}
 (Elmsford:\ Pergamon)
\item[]
Welch\ G and Bishop\ G 2002 An introduction to Kalman filter 
({\it http://www.cs.unc. edu/$\sim$welch/kalman/})
\item[] 
Zech G  2002 Frequentist and Bayesian confidence limits 
    {\it Eur. Phys. J. direct} {\bf C12} 1 
\item[] 
Zellner A 1986 Bayesian solution to a problem posed by Efron
{\it Am. Stat.} {\bf 40} 330
\end{itemize}

\end{document}